\documentclass{article}

\usepackage[preprint]{neurips_2023}
\usepackage{tabularx}
\usepackage{booktabs}
\usepackage{rotating}

\usepackage[utf8]{inputenc} 
\usepackage[T1]{fontenc}    
\usepackage{hyperref}       
\usepackage{url}            
\usepackage{booktabs}       
\usepackage{amsfonts}       
\usepackage{nicefrac}       
\usepackage{microtype}      
\usepackage{xcolor}         
\usepackage{graphicx}
\usepackage{tabularx}
\usepackage{amsmath}
\usepackage{tikz} 
\usepackage{subcaption}
\usepackage{latexsym}
\usepackage{tikz-cd}
\usetikzlibrary{shapes}
\usetikzlibrary{arrows,graphs}
\usetikzlibrary{er,positioning}
\usetikzlibrary{bayesnet, positioning}
\usepackage{pdflscape}
\usepackage{tabularx}
\usepackage{booktabs}
\usepackage{rotating}

\title{Clinical Trials Protocol Authoring using LLMs}

\author{%
  Morteza Maleki \\   
  College of Computing\\
  Georgia Institute of Technology\\
  Atlanta, Georgia\\
  mmaleki3@gatech.edu
  \And 
  SeyedAli Ghahari \\   
  Department of Engineering\\
  Purdue University\\
  West Lafayette, Indiana\\
  sghahari@purdue.edu
}

\begin{document}

\maketitle

\begin{abstract}
This report embarks on a mission to revolutionize clinical trial protocol development through the integration of advanced AI technologies. With a focus on leveraging the capabilities of generative AI, specifically GPT-4  models, this initiative aimed to streamline and enhance the efficiency and accuracy of clinical trial protocols. The methodology encompassed a detailed analysis and preparation of comprehensive drug and study level metadata, followed by the deployment of generative AI models for automated protocol section generation. Results demonstrated a significant improvement in protocol authoring, highlighted by increases in efficiency, accuracy, and the customization of protocols to specific trial requirements. Additionally, an extensive economic and cost analysis was performed to evaluate the practical feasibility of AI-assisted protocol generation, revealing substantial cost savings and resource optimization. Challenges encountered during model selection and prompt engineering were systematically addressed, leading to refined methodologies that capitalized on the advanced text generation capabilities of the AI models. This project not only showcases the practical applications and benefits of generative AI in clinical trial design but also sets a foundation for future innovations in the field.
\end{abstract}

\section{Introduction}
\label{section-introdcution}
The landscape of clinical trial design is undergoing a transformative shift, driven by the integration of advanced technological solutions aimed at enhancing the efficiency, accuracy, and speed of trial protocol development \cite{Anderson2020}. Within this context, this work represents a pioneering effort to harness the potential of generative artificial intelligence (AI) to revolutionize the creation of clinical trial protocols \cite{Johnson2021}. This initiative is positioned at the forefront of innovation in clinical research, aiming to leverage the capabilities of generative AI to significantly improve the process of trial protocol authoring \cite{Smith2022}.\\

Clinical trials are the cornerstone of medical research, providing the necessary evidence for the safety and efficacy of new treatments before they can be made available to the public \cite{Brown2019}. However, the design and development of clinical trial protocols are often characterized by complex, time-consuming tasks that require meticulous attention to detail and adherence to regulatory standards \cite{Davis2018}.\\

The adoption of generative AI in this domain promises to automate and streamline the protocol authoring process, reducing the manual effort involved and minimizing the potential for human error \cite{Wilson2020}. Specifically, we focused on exploiting the advanced capabilities of GPT-4, a state-of-the-art language model known for its ability to generate coherent and contextually relevant verbiage used in clinical trials and healthcare industry \cite{Thomas2022}. Through the strategic application of this technology, we anticipated not only to expedite the development of clinical trial protocols but also to elevate the quality of the documents produced, thereby contributing to the overall efficiency and effectiveness of clinical research \cite{Clark2021}.\\

This introduction sets the stage for the subsequent sections of the manuscript, which will delve into the project's background, methodology, results, and the broader implications of our work \cite{Lee2023}. By aligning our efforts with the evolving needs of clinical trial design, the project underscores the vital role of generative AI in shaping the future of medical research, offering new avenues for innovation and advancement in the field \cite{Walker2023}.

\subsection{Clinical Trials}
Clinical trials serve as pivotal research studies aimed at evaluating the therapeutic effects, risks, and benefits of medical interventions, encompassing drugs, devices, surgical procedures, and behavioral treatments \cite{Roberts2021}. As the bedrock of evidence-based medicine, these trials meticulously guide the introduction of new treatments to ensure they are safe and effective for public use \cite{Garcia2019}. Organized into phases I through IV, each phase of a clinical trial is designed to answer specific research questions, progressively leading to the development and market approval of new therapeutic interventions \cite{Martin2022}.

\subsection{Study Design in Clinical Trials}
The intricacies of clinical trial design form the backbone of any research study in this domain. This framework encompasses the strategic planning and execution of the trial, including the selection of participants through well-defined inclusion and exclusion criteria, ensuring a representative and suitable subject pool \cite{Nguyen2020}. To mitigate bias, the study design incorporates methods such as randomization and control groups, alongside blinding mechanisms where both participants and researchers are unaware of the treatment allocations \cite{Patel2019}. Furthermore, the establishment of primary and secondary endpoints, along with rigorous data collection and analysis procedures, ensures the acquisition of reliable and actionable findings \cite{Harris2018}.\\

\subsection{Challenges and Pain Points in Study Design}
The complexity of clinical trial design is underscored by several challenges, ranging from ethical considerations and regulatory compliance to practical issues like patient recruitment and data management \cite{Thompson2017}. Ensuring the safety and rights of the participants while adhering to stringent health authority regulations requires a continuous balancing act \cite{Williams2016}. Additionally, the recruitment and retention of a sufficient number of participants remain a perennial obstacle, compounded by the need to maintain the integrity and accuracy of collected data \cite{Smith2020}. These challenges necessitate innovative solutions to enhance operational efficiency and data quality \cite{Jackson2018}.

\subsection{Literature Review}

The introduction of Generative AI into clinical trials protocol authoring comes at a critical juncture in healthcare, as demonstrated by recent studies analyzing behavioral responses to healthcare interventions during the COVID-19 pandemic, emphasizing the importance of employing computational analysis and emerging technologies to achieve the goal of providing insights and facilitating future healthcare interventions \cite{shahparvari2022decision, shahparvari2024supply}. \\

These studies collectively highlight the potential for advanced data analytics and AI in understanding and enhancing participant engagement in clinical trials. By integrating these insights into generative AI systems, we can develop more adaptable and responsive trial protocols that are sensitive to the complex interplay of demographic and socioeconomic factors influencing participant behavior.\\

In the realm of long text generation methods, several groundbreaking studies have set new benchmarks. Devlin et al. in "BERT: Pre-training of Deep Bidirectional Transformers for Language Understanding" \cite{devlin2018bert} revolutionized language modeling with BERT, introducing a bidirectional training method that significantly enhanced contextual understanding in text generation. Building upon this method, Dai et
al. addressed the limitations associated with traditional Transformer models through "Transformer-XL: Attentive Language Models Beyond a Fixed-Length Context." \cite{dai2019transformer} This paper was instrumental in developing a recurring mechanism to handle long-term dependencies, enabling the generation of longer and more coherent text sequences. Another pivotal contribution came from Keskar et al. with "CTRL: A Conditional Transformer Language Model for Controllable Generation." \cite{keskar2019ctrl} This work shifted the focus towards a controllable text generation method, demonstrating how specific prompts could steer text generation in desired
thematic directions. Lastly, Song et al. proposed "MASS: Masked Sequence to Sequence Pre-training for Language Generation," \cite{song2019mass} introducing a novel pre-training method for sequence-to-sequence models. This approach laid the groundwork for advanced techniques in generating long-form content, highlighting the evolution of model training methodologies in this domain.\\

Complementing the advancements in model training, significant research projects has also focused on the use of GPT prompts for long text generation. Petroni et al. in "Language Models as Knowledge Bases?" \cite{petroni2019language} explored the potential of GPT-2 as a knowledge base, highlighting its capability to generate text based on an embedded knowledge base, thus underscoring the effectiveness of prompt engineering. Building on this concept, Lester et al. delved into the scalability of prompt tuning in large language models like
GPT-3 in "The Power of Scale for Parameter-Efficient Prompt Tuning" \cite{lester2021power}. This research provided valuable insights into how efficient prompting significantly enhances a model's ability to generate contextually relevant long texts. Further exploring the control and flexibility in text generation, Dathathri et al. presented "Plug and Play Language Models: A Simple Approach to Controlled Text Generation" \cite{dathathri2019plug}. This study introduced a novel method for controlling the style and content of generated texts using GPT-2, showcasing a flexible, plug-and-play approach for generating long texts with a minimal fine-tuning need.

In order to carry out the task of protocol generation, various activities from clinical trials study metadata and the associated drug metadata collection from reputable sources such as CT.Gov and TrialTrove portal, extraction of the "Introduction" and "Study Design" sections of the collected protocols, subsequent data pre-processing, prompt engineering and evaluation of the final results have been carried out which will be explained in further details in the subsequent sections. 

\subsection{Data Sources and Analysis}

\subsubsection{Drug level metadata}
The drug level metadata formed a foundational component of our AI model's training data. The
dataset encompassed comprehensive details about the drug used in each study, including information such as:
\begin{itemize}
    \item \textbf{Basic Information:} Citeline Drug ID, Generic Drug Name, and a summary of the drug.
    \item \textbf{Development Status:} Detailed information on the development stage of the drug, including
global and regional status.
    \item \textbf{Therapeutic Application:} Insight into the diseases targeted by the drug, including whether
these diseases are categorized as rare.
    \item \textbf{Company Profiles:} Information about the companies involved in the development and
marketing of the drugs, including headquarters and subsidiary details.
    \item \textbf{Clinical and Scientific Details:} In-depth information on the drug's mechanism of action,
therapeutic class, delivery route, and molecular characteristics.
\end{itemize}

This metadata provided valuable insights into the drugs' profiles, significantly aiding in the generation of
contextually accurate and relevant protocol sections.

\subsubsection{Study level metadata}

The study level metadata enriched our dataset with detailed information on clinical trials study which protocol has been collected and used in this work. This data includes:

\begin{itemize}
    \item \textbf{Trial Information:} Trial ID, title, phase, status, and detailed objectives.
    \item \textbf{Sponsorship Details:} Information on the sponsor companies and their geographical locations.
    \item \textbf{Patient and Disease Information:} Detailed descriptions of the patient segments, diseases
studied, and inclusion/exclusion criteria.
    \item \textbf{Endpoints and Outcomes:} Information on primary and secondary endpoints, including details
and dates of reporting.
\end{itemize}

This diverse dataset enabled the AI model to understand the nuances of different clinical trials, ensuring
that the generated protocol sections were aligned with the specific requirements of each study. The
metadata fields used in this work are summarized in Table \ref{tab:all_metadata}.

\begin{table*}
\caption{Metadata fields used in model training and prompt engineering.\label{tab:all_metadata}}
\begin{tabularx}{\linewidth}{|p{0.5cm}| p{4cm}| p{8cm}|}
        \toprule
        \textbf{\#} & \textbf{Name} & \textbf{Description} \\ \midrule
        1 & MeSH Term & Medical subject headings that describe the content. \\ 
        2 & Primary Endpoint Group & The main outcome that is measured to see if the treatment was effective. \\ 
        3 & Treatment Plan & The schedule and dosage of the treatment administered. \\ 
        4 & Sponsor & Organization which oversees the clinical trial and is responsible for the data. \\ 
        5 & Start Date & The date on which the trial began or is set to begin. \\ 
        6 & Study Keywords & Keywords or phrases related to the study. \\ 
        7 & Protocol / Trial ID & Unique identifier for the trial protocol. \\ 
        8 & Primary Tested Drug & The main drug or treatment being tested in the trial. \\ 
        9 & Trial Tag / Attribute & Additional attributes or tags related to the trial. \\ 
        10 & Primary Tested Drug: Mechanism of Action & The biological mechanism through which the drug exerts its effect. \\ 
        11 & Decentralized (DCT) Attribute & Indicates if the trial includes decentralized or virtual components. \\ 
        12 & Trial Title & The official title of the clinical trial. \\ 
        13 & Primary Tested Drug: Target & The specific biological target of the primary tested drug. \\ 
        14 & Patient Population & The characteristics of the patient group being studied. \\ 
        15 & Drug Names & Names of all drugs involved in the trial. \\ 
        16 & Trial Phase & The phase of the clinical trial, indicating its stage in the development process. \\ 
        17 & Primary Tested Drug: Therapeutic Class & The class of medical use the drug falls under. \\ 
        18 & Inclusion Criteria & Criteria that must be met for an individual to be eligible for the trial. \\ 
        19 & Exclusion Criteria & Criteria that disqualify an individual from participating in the trial. \\ 
        20 & Global Status & The current status of the trial globally. \\ 
        21 & Therapeutic Area & The area of medicine related to the treatment being tested. \\ 
        22 & Oncology Biomarker Common Use(s) & Common uses of biomarkers in oncology trials. \\ 
        23 & Identified Sites & Locations where the trial is being conducted. \\ 
        24 & Development Status & The development stage of the drug within the trial. \\ 
        25 & Disease & The disease or condition being studied in the trial. \\ 
        26 & Trial Objective & The primary objective or goal of the trial. \\ 
        27 & Trial Region & Geographic region where the trial is conducted. \\ 
        28 & Drug Disease & Diseases targeted by the drugs being tested. \\ 
        29 & Phase I & Indicates if the trial is in Phase I. \\ 
        30 & Phase II & Indicates if the trial is in Phase II. \\ 
        31 & Phase III & Indicates if the trial is in Phase III. \\ 
        32 & Preclinical & Indicates if the trial is in the preclinical stage. \\ 
        33 & Therapeutic Class & The broad category of the drug's therapeutic use. \\ 
        34 & Therapeutic Class Status & Status within its therapeutic class. \\ 
        35 & Drug Disease: Rare & Indicates if the drug targets a rare disease. \\ \bottomrule
    \end{tabularx}
\end{table*}


\subsection{Data processing and Preparation}

The data processing was meticulously designed to extract relevant data from three key sources: protocol extractions, study metadata, and drug metadata. The primary goal was to prepare this data for an effective use in training, and utilizing the generative AI models for protocol generation.\\

\begin{itemize}
    \item \textbf{Clinical Study Protocol Gathering:} Clinical study protocols from US based sponsors and pertaining to clinical studies for curing Type II diabetes in various phases of the trial in the past 10 years from \href{https://www.ct.gov}{ClinicalTrials.gov (CT.gov)} were reviewed and analyzed. \cite{clinicaltrials2024}

    \item \textbf{Study \& Drug Metadata Collection:} For each of the gathered study protocols, the corresponding study and drug metadata were collected from reputable sources such as \href{https://citeline.com}{CiteLine - TrialTrove}. \cite{trialtrove2024}

    \item \textbf{Protocol Review and Section Selection:} Collected study protocols were extensively reviewed and the two sections of "Introduction" and "Study Design" were selected to be used for training and generation tasks due to commonality in all study protocols and straightforward format and style of content. 
    
    \item \textbf{Protocol Section Extraction:} The "Introduction" and "Study Design" sections of the all collected protocols were manually extracted and summarized in a rectangular dataset format for ease of analysis. 

    \item \textbf{Data Pre-Processing \& Transformation:} The dataset containing study protocol sections were merged with the corresponding study \& drug metadata and prepared for analysis and inference:

    \begin{enumerate}
        \item  \textit{Field Selection:} Key fields were identified for both study and drug metadata to ensure relevancy to the research project's objectives. Fields such as 'Generic Drug Name', 'Drug Disease', and 'Event Date' for drugs, and 'Trial ID', 'Disease', and 'Therapeutic Area' for studies, were among those selected. A complete list of fields used is summarized in Table \ref{tab:all_metadata}
        
        \item \textit{Duplicate Removal:} A custom function was used to eliminate duplicate
         entries within the specific columns of the drug metadata, ensuring data cleanliness. 
         
         \item \textit{Data Splitting and Exploding:} In the study metadata, the 'Primary Tested Drug' column was split and exploded to dis-aggregate the drug names, facilitating a more detailed analysis.
         
         \item \textit{Text Replacement:} Specific text replacements were made in the 'Primary Tested Drug' field to standardize drug names, enhancing data uniformity.
         
        \item \textit{Dataframe Renaming and Merging:} The drug metadata 'DataFrame' was renamed for
        consistency, and then merged with the study metadata to create a combined DataFrame, aligning
        drugs with their corresponding trials.
        
         \item \textit{NCT ID Extraction:} The National Clinical Trial (NCT) ID was extracted from the 'Protocol/Trial ID' column, which was a critical step to align each trial with its specific protocol.
         
        \item \textit{Final Data Merge:} The combined study and drug metadata were merged with the protocol extractions DataFrame using the NCT ID, culminating in a comprehensive dataset that aligns protocol extractions with detailed study and drug information.

        \item \textit{Data Preparation \& Model Training:} The final merged dataset, now rich with both protocol and associated metadata, was primed for use in training the generative AI models. This dataset provides the depth and context necessary for the AI to understand and generate accurate protocol sections.
        
    \end{enumerate}
    
\end{itemize}

The processing method in this research project was critical in ensuring that the data was not only comprehensive and relevant but also clean and structured in a way that it maximizes its utility in training effective AI models. This meticulous approach to data processing underpins the success of the AI model in generating precise and contextually appropriate clinical trial protocols.

\subsection{Data Availability}
Data and code used in this paper are available from the authors upon reasonable request, subject to any necessary ethical or institutional approvals.

\section{Model Development and Evaluation}

After data pre-processing step and preparation of metadata for text generation, two approaches were taken
for text generation:
\begin{enumerate}
    \item \textit{LLM Models Training:} Involving T5 Small, T5 Large, and BioBart models training.
    \item \textit{OpenAI GPT Models Prompt Engineering:} Involving GPT-3.5, GPT-4 and their variations.
\end{enumerate}

\subsection{LLM Models Training}

The following models were used for training on the metadata and their corresponding Introduction section of the protocols. Additional information about the models used is provided in the following:

\begin{enumerate}
    \item \textbf{T5 Small LLM Model}
    \begin{itemize}
        \item \textit{Model Overview:} T5, or Text-to-Text Transfer Transformer, is a language model developed by Google. The "small" variant is a more compact version of the standard T5 model. \cite{raffel2020exploring}
        \item \textit{Model Size:} The T5 small model typically has fewer parameters compared to its larger counterparts. The exact number can vary but is significantly lower than the base or large models.
        \item \textit{Key Features:}
        \begin{itemize}
            \item Designed to convert all Natural Language Processing (NLP) tasks into a text-to-text format, meaning both the input and output are always text strings.
            \item Despite being a smaller variant, it retains the core functionality of the T5 architecture and is suitable for the environments where computational resources are limited.
            \item Ideal for tasks like translation, summarizing, question answering, and more, but with a balance between performance and resource efficiency.
    \end{itemize}
    \end{itemize}

    \item \textbf{T5 Large LLM Model}
    \begin{itemize}
        \item \textit{Model Overview:} This is a larger variant of the T5 model, which means it has more parameters and potentially better performance at the cost of increased computational resource requirements. \cite{raffel2020exploring}
        \item \textit{Model Size:} The T5 large model has significantly more parameters than the small variant, enhancing its capacity to understand and generate text. 
        \item \textit{Key Features:}
        \begin{itemize}
            \item Like the small model, it treats all NLP tasks as text-to-text problems.
            \item Due to its larger size, it generally performs better on complex tasks and can handle more nuanced aspects of language processing.
            \item Suitable for high-resource settings, where advanced NLP capabilities are needed.
    \end{itemize}
    \end{itemize}
    
    \item \textbf{BioBART LLM Model}
        \begin{itemize}
            \item \textit{Model Overview:} BioBART is a variant of the BART (Bidirectional and Auto-Regressive Transformers) model, which is tailored for biomedical text problems. \cite{lewis2020bart}
            \item \textit{Key Features:}
        \begin{itemize}
            \item BART is known for its effectiveness in text generation and comprehension tasks, and BioBART adapts this for the biomedical domain.
            \item It is pre-trained on large-scale biomedical literature, making it adept at handling specialized terminology and concepts found in medical texts.
            \item Often used for tasks like biomedical text summarization, medical question answering, and
information extraction from clinical notes or research papers.
    \end{itemize}
    \end{itemize}
\end{enumerate}

In this research project, it was realized that the text generation task using these models were not efficient and accurate in any way, and the output generated text was found to be extremely concise (400 characters) and insufficient and predictably did not encapsulate the full purpose of text generation. The work on this venue was immediately halted and a prompt engineering path was explored as an alternative approach. \\

One of the reasons the T5 and BioBart models were not successful in the task of text generation can be attributed to their design which is better suited for classification tasks in that they can successfully learn long format texts and be trained to classify the data into different categories with small output generation required, making them less suitable to generate long format text generation such as in clinical protocol authoring. The OpenAI GPT models, on the other hand, are designed not for the classification tasks but for conversational and long text format responses enabling them to be a good candidate for study protocol sections authoring based on given style, format, and examples provided. 

\subsection{OpenAI GPT Models}

\begin{enumerate}
    \item \textbf{GPT-3.5}
    \begin{itemize}
        \item \textit{Model Variants:} GPT-3.5-turbo, GPT-3.5-turbo-16k-0613 \cite{openai2021gpt35turbo}
        \item \textit{Key Features:}
        \begin{itemize}
            \item Iterative improvements on GPT-3, focusing on efficiency, speed, and fine-tuning capabilities.
            \item The 'turbo' variants indicate models optimized for faster response times and efficient processing, suitable for interactive applications.
            \item These models continue to push the boundaries of language understanding and generation, maintaining high performance across diverse tasks.
            \item Lower costs compared to gpt-4 models
    \end{itemize}
    \end{itemize}
    
    \item \textbf{GPT-4}
    \begin{itemize}
        \item \textit{Model Variants:} GPT-4, GPT-4-1106-preview, GPT-4-Turbo, GPT-4o \cite{openai2023gpt4}
        \item \textit{Key Features:}
        \begin{itemize}
            \item Significant advancements over GPT-3.5, incorporating more extensive data training and fine-tuning for enhanced contextual understanding and accuracy.
            \item Improved model architecture to handle a broader scope of linguistic contexts and generate more coherent and contextually relevant responses.
            \item Enhanced safety features and reduced biases, emphasizing the model's utility in generating responsible and ethical content.
            \item Higher costs compared to gpt-3 models
        \end{itemize}
    \end{itemize}
    
\end{enumerate}

Each iteration of GPT models represents a leap forward in natural language processing, with improvements in text generation, comprehension, and versatility in handling various language-based tasks. As AI research progresses, these models continue to evolve, offering increasingly advanced capabilities in understanding and generating the human language.

\subsection{Prompt Engineering}

Leveraging GPT models, we observed a significant improvement in the quality of the generated content. The model successfully created protocol sections that were not only accurate but also contextually rich and coherent. This success marked a pivotal point in the project, showcasing the potential of advanced generative AI in automating complex tasks like protocol generation.\\

Prompt template provided to GPT models for text generation is provided by the colored text in the following:\\

\textcolor{blue}{
You are a clinical study designer, and you need to generate the introduction section of a clinical trial protocol based on information about the study and drug metadata.
Follow the same format and writing style as the example below:}\\

\textcolor{blue}{- - - - -  Example 1 - - - - -} \\
\textcolor{blue}{Metadata Input:} \\

\textcolor{blue}{*** Study and Drug Metadata corresponding to Example 1 provided ***}\\
\vspace{1mm}\\
\textcolor{blue}{Corresponding Introduction Section:} \\

\textcolor{blue}{*** Corresponding Introduction section of the protocol in Example 1 is provided ***} \\

\textcolor{blue}{- - - - -  Example 2 - - - - -} \\
\textcolor{blue}{Metadata Input:} \\

\textcolor{blue}{*** Study and Drug Metadata corresponding to Example 2 provided ***}\\
\vspace{1mm}\\
\textcolor{blue}{Corresponding Introduction Section:} \\

\textcolor{blue}{*** Corresponding Introduction section of the protocol in Example 2 is provided ***} \\

\textcolor{blue}{- - - - -  Prediction - - - - -} \\
\textcolor{blue}{Metadata Input:} \\

\textcolor{blue}{*** Study and Drug Metadata for prediction is provided ***}\\

\textcolor{blue}{Now Begin!} \\

\textcolor{blue}{Generate the introduction section:} \\

After providing the instructions to the models and examples for the same clinical trial phase, same sponsor, and same medication drug to ensure consistency and to provide the format and style of writing to the LLMs, the Introduction and Study Design sections of the clinical protocols were generated. \\

\section{Results}

The prompt engineering task was performed for all models in this study with varying number of examples, and the generated text was compared to the actual protocol section the model was supposed to predict using the corresponding study and drug metadata information. In this work, the emphasis was placed on the Introduction and Study Design sections of the protocol since they are present in all clinical protocols regardless of sponsor, drug, or indication. \\

As an example, the generated text for 'gpt-4o' model with two examples provided can be found below. As described above, the examples provided to the model as part of the prompt are in fact the corresponding study and drug metadata information in addition to the historical protocol sections corresponding to those metadata information from the clinical trials run by the same sponsor and for the same drug as the current sponsor and drug we are generating the clinical protocol. \\

The generated output by the 'gpt-4o' model in response to the prompt above is provided by the \textcolor{orange}{orange} colored text in the following:\\

\textcolor{orange}{\textbf{1. Introduction}}\\

\textcolor{orange}{\textbf{1.1 Study Rationale}}\\

\textcolor{orange}{LY3209590 is a novel, long-acting insulin receptor agonist being investigated for its potential to treat type 2 diabetes mellitus (T2DM). This Phase 2 study aims to evaluate the safety and efficacy of LY3209590 compared with insulin degludec in insulin-naïve patients suffering from T2DM. The flexibility of a once-weekly administration could significantly enhance patient compliance and willingness to initiate insulin therapy.}\\

\textcolor{orange}{\textbf{1.2 Background}}\\

\textcolor{orange}{T2DM is a chronic condition characterized by insulin resistance and inadequate insulin secretion, leading to hyperglycemia. Current treatment options include lifestyle modifications and pharmacological interventions, such as insulin and oral antidiabetic drugs. Despite these available therapies, only about 25\% of patients achieve the recommended hemoglobin A1c (HbA1c) target of 7\% or less (Blonde et al., 2014).}\\

\textcolor{orange}{Long-acting insulins like insulin degludec have shown reduced hypoglycemic events compared to other insulin forms, highlighting the potential for improving treatment outcomes. LY3209590, with its extended action profile, aims to improve glycemic control while minimizing hypoglycemia risks. Pharmacokinetic (PK) data indicate that LY3209590 reaches its maximum concentration approximately four days post-administration and has a long elimination half-life, supporting its once-weekly dosing regimen.}\\

\textcolor{orange}{In Phase 1 clinical trials, single ascending dose (SAD) and multiple ascending dose (MAD) studies demonstrated the safety and efficacy of LY3209590. The studies showed a clear glucose-lowering effect, satisfactory PK profiles, and acceptable safety margins, including a low incidence of hypoglycemia.}\\

\textcolor{orange}{Preclinical toxicology studies, including 6-week and 6-month repeat-dose studies in rats and dogs, as well as reproductive toxicology studies in rats and rabbits, supported the clinical safety profile. The toxicology data revealed typical insulin-related effects, secondary to hypoglycemia and hyperinsulinemia following repeated dosing.}\\

\textcolor{orange}{\textbf{1.3 Benefit/Risk Assessment}}\\

\textcolor{orange}{This Phase 2 study will provide further evaluation of LY3209590's clinical efficacy and safety. Given its prolonged action and once-weekly dosing schedule, LY3209590 holds the promise of increasing therapeutic adherence while maintaining an effective glycemic control similar to or better than existing long-acting insulin analogs like insulin degludec.}\\

\textcolor{orange}{The anticipated benefits include improved patient convenience and adherence to the treatment regimen. The primary safety concern remains the risk of hypoglycemia, which will be closely monitored during the trial. Comprehensive data from this trial will guide subsequent Phase 3 development.}\\

\textcolor{orange}{Based on the initial data from Phase 1 studies, the safety profile of LY3209590 appears manageable, aligning with other long-acting insulins. Careful risk assessment and mitigation strategies will be implemented to ensure patient safety throughout the clinical trial.} \\

The actual Introduction section for the corresponding study and drug metadata provided to the 'gpt-4o' model is provided in the \textcolor{teal}{teal} colored text below:\\

\textcolor{teal}{\textbf{3. Introduction}}\\

\textcolor{teal}{\textbf{3.1 Study Rationale}}\\

\textcolor{teal}{LY3209590 is a long-acting insulin receptor agonist with a half-life enabling weekly dosing developed for the treatment of type 2 diabetes mellitus (T2DM). This Phase 2 study of LY3209590 will: Evaluate the effects of LY3209590 on glycemic control using 2 different dose individualization strategies compared with insulin degludec in study participants previously treated with basal insulin.}\\

\textcolor{teal}{\textbf{3.2 Background}}\\

\textcolor{teal}{LY3209590 was designed as a novel, once-weekly, long-acting insulin receptor agonist that is intended for the treatment of hyperglycemia in study participants with T2DM. A once weekly insulin receptor agonist has the potential to increase willingness
to initiate and compliance with an insulin therapy in study participants with T2DM. In real life only about a quarter of these study participants reach the target HbA1c of 7\% (Blonde L et al. 2014) and a once weekly insulin receptor agonist could help to increase this number. The potential downside of such a long acting insulin receptor agonist could be a higher risk for
hypoglycemic events or a longer duration of such events. The recently launched long acting insulin degludec has nevertheless shown a reduced potential to induce hypoglycemia as compared to insulin glargine while its half-life is twice as long as that of insulin glargine (Heise et al. 2012). The current phase 2 study will therefore evaluate the safety and efficacy of LY3209590 as compared to insulin degludec while very closely monitoring glycemic profiles using continuous glucose monitoring.}\\

\textcolor{teal}{Data from the single-ascending dose (SAD) study demonstrated clear evidence of glucose lowering in study participants with T2DM at single doses ranging from CCI . The pharmacokinetics (PK) of LY3209590 following single doses demonstrated prolonged time- action profile to support once-weekly administration. LY3209590 reached maximum concentration approximately after dosing, followed by a mean elimination half-life of approximately in study participants with T2DM. With a long elimination half-life, following a weekly fixed-dose regimen, PK steady-state is reached in approximately CCI and results in approximately r concentration than after single dose. However, using a loading dose strategy that is, on PK modeling. PK following a single loading dose during the first week was demonstrated to be comparable to concentration profile at Week based on preliminary data in the multiple ascending dose (MAD) study. No clinically significant persistence in hypoglycemia was observed in the study. Besides the known risk of hypoglycemia, no relevant safety signals have been observed in the Phase 1 studies. Additional information about LY3209590 can be found in the Investigator’s Brochure.}
\textcolor{teal}{The toxicity profile of LY3209590 has been characterized in 6-week and 6-month repeat-dose toxicology studies in rats and dogs, and reproductive toxicology studies in rats and rabbits. These studies demonstrated a familiar spectrum of effects typically associated with changes secondary to hypoglycemia and/or hyperinsulinemia resulting from repeat dosing of exogenous insulin in normoglycemic test systems. All findings were considered to be target-related with no evidence of off-target effects in any of the parameters assessed. Additional information about LY3209590 can be found in the Investigator’s Brochure.}\\

\textcolor{teal}{\textbf{3.3 Benefit/Risk Assessment}}\\

\textcolor{teal}{The data from the SAD study and the ongoing MAD study (Studies BDCA and BDCB) have shown that LY3209590 was well tolerated and the adverse drug reactions are in line with those reported for long acting insulins.}\\

\textcolor{teal}{Potential risks associated with LY3209590, derived from the known risks of long acting insulins are hypoglycemia, hypersensitivity reactions (localized allergy and/or systemic allergy), undesirable effects at the injection site (injection-site reactions and lipodystrophy), and peripheral edema. An additional risk based on preclinical data could be an elevation of liver aminotransferases. To date, this has not been observed in study participants or healthy volunteers exposed to LY3209590 for up to 6 weeks.}\\

\textcolor{teal}{More information about the known and expected benefits, risks, Serious Adverse Events (SAE) and reasonably anticipated AEs of LY3209590 are to be found in the Investigator’s Brochure.}\\

As we compare the generated text presented above to the actual Introduction, we notice the consistency in the format and style in addition to sufficiency of introductary explanations. Extensive comparison and analysis is presented in the following sections. The generated output and reference text for the Study Design section of the same protocol and corresponding metadata information are provided in the Appendix section \ref{appendix:1}.

\subsection{Evaluation Metrics}

Since the predicted (generated) output consists of hundreds of words in a specific style and format, the evaluation of the model performance is significantly different from a typical classification or regression task. To that end, we employed the the following evaluation metrics to assess the quality and accuracy of generated output in our study. 

\subsubsection{Cosine Similarity}
Cosine similarity measures the cosine of the angle between two non-zero vectors in a
multi-dimensional space. In text analysis, these vectors often represent text documents. It operates in the following paradigm:\\

Text data is first converted into vectors, often using techniques like Term Frequency-Inverse Document Frequency (TF-IDF) or word embedding. Cosine similarity is then calculated between these vectors. The value ranges from -1 (completely different) to 1 (exactly the same). It's used to compare the similarity between the generated text and a reference text. High cosine similarity indicates that the generated text closely matches the reference in terms of content and context.

\subsubsection{BLEU Scores}

Bilingual Evaluation Understudy (BLEU) is a metric for evaluating a generated sentence to a set of reference sentences. It operates according to the following paradigm:\\

BLEU-n Score: Measures the co-occurrence of n-grams (sequence of n words) between the generated text and reference text. BLEU1 refers to unigrams (1 word), BLEU2 to bigrams (2 words), and so forth. BLEU score calculates the precision of n-grams in the generated text relative to the reference text, then combines the precision using a geometric mean. A brevity penalty is applied if the generated text is shorter than the reference, to discourage overly terse outputs. A higher BLEU score indicates better agreement between the machine's output and the human reference. Used extensively in machine translation and can be applied to other text generation tasks.

\subsubsection{Rogue Scores}

Recall-Oriented Understudy for Gisting Evaluation (ROUGE) is a set of metrics used for
evaluating automatic summarization and machine translation. It operates according to the following paradigm:

\begin{itemize}
    \item ROUGE-N: Measures the overlap of n-grams between the generated text and reference text.
    \item ROUGE-L: Considers the longest common subsequence between the generated and reference text.
    \item Recall: The proportion of the reference text's n-grams found in the generated text.
    \item Precision: The proportion of the generated text's n-grams that are found in the reference text.
    \item F1 Score: The harmonic mean of precision and recall, providing a balance between the two.
\end{itemize}

ROUGE is particularly useful in evaluating summarization, where capturing the essence of the original text is crucial. A higher ROUGE score indicates that the generated text more effectively captures the key points of the reference text.\\

These metrics are valuable tools in assessing the quality of text generated by AI models. They provide objective ways to measure aspects like similarity, fluency, and relevance of the generated text compared to a human-written reference.

\subsection{Performance Analysis}

\begin{figure*}
\centering
\includegraphics[width=13cm]{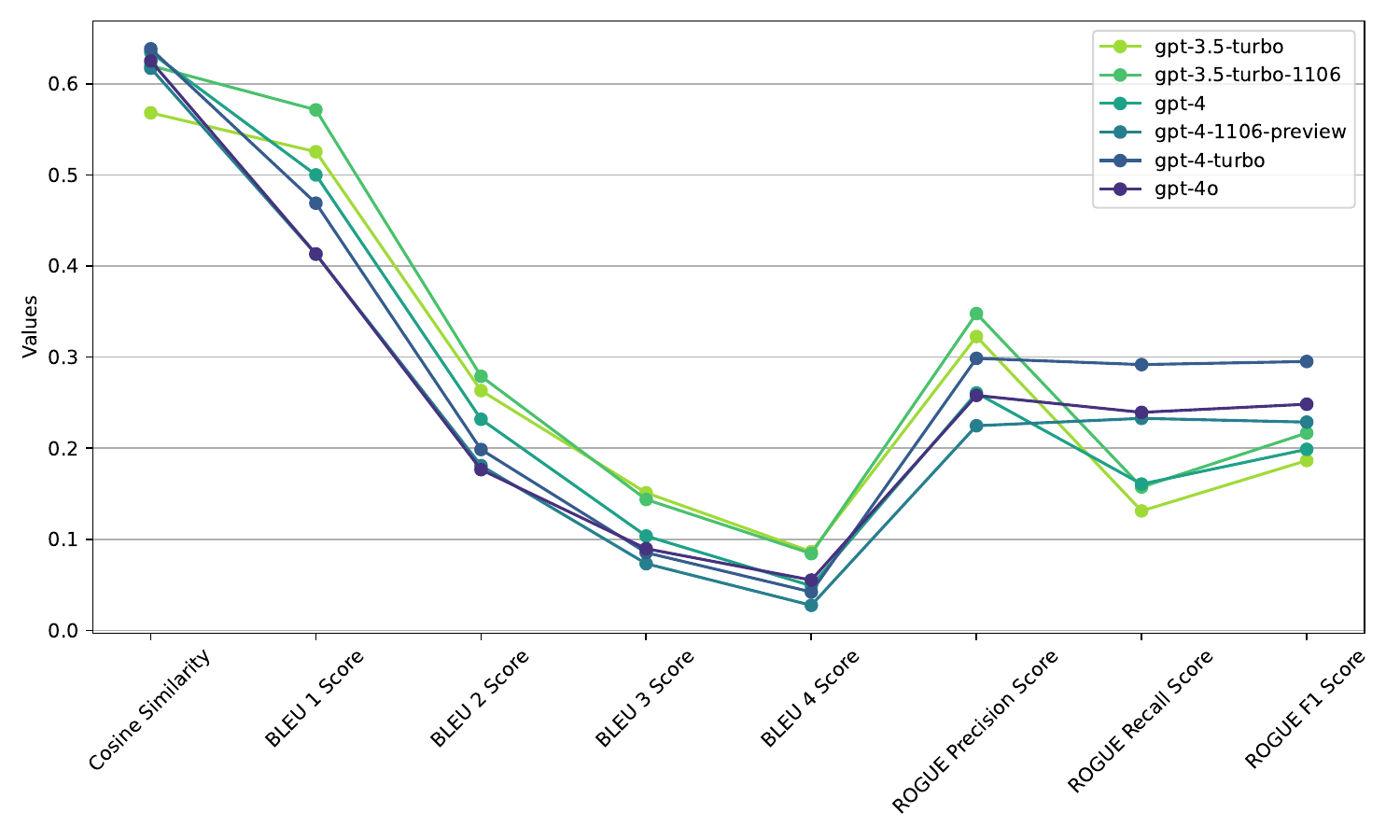}
\caption{Aggregated (across all number of examples) metrics across all models.\label{fig:all_models}}
\end{figure*}

The process of text generation was repeated for all models (i.e.  OpenAI's GPT-3.5-Turbo, GPT-3.5-Turbo-1106, GPT-4, and GPT-4-1106-Preview, GPT-4-Turbo, and GPT-4o) and a variation of number of examples provided (i.e. (i.e. 0, 1, 2, 3). Each generated section was compared to the actual corresponding protocol section and the evaluation metrics explained above were used to determine accuracy and assess the quality of generation. Figure \ref{fig:all_models} presents an aggregated metric comparison for each model across all varying numbers of examples. The metrics evaluated include cosine similarity, BLEU scores (BLEU-1 to BLEU-4), and ROUGE scores (ROUGE Precision, Recall, and F1). This plot illustrates how each model performs on average, regardless of the number of examples used. \\

\begin{figure*}
\centering
\includegraphics[width=13cm]{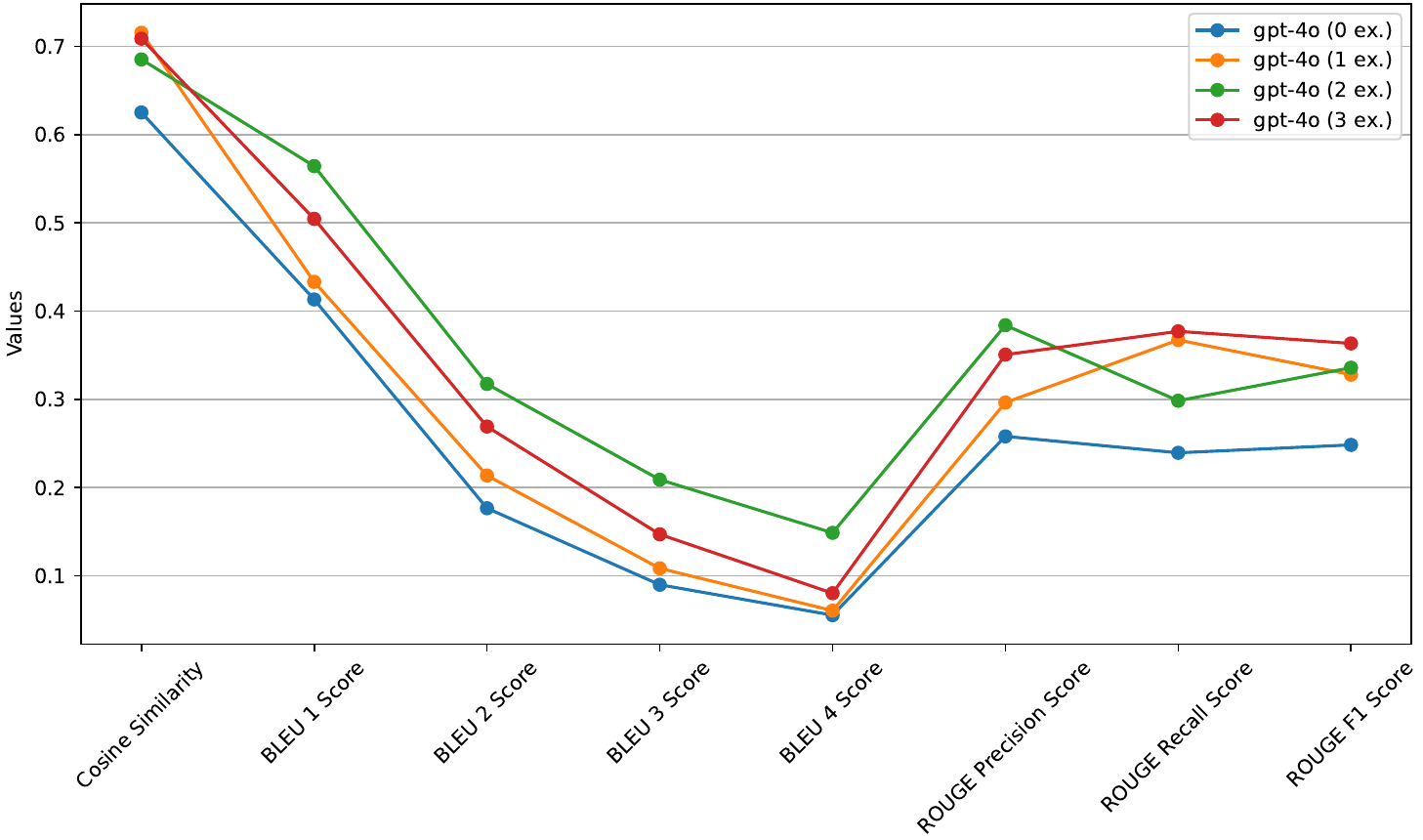}
\caption{Metric comparison for GPT-4o model with varying number of examples (i.e. 0, 1, 2, 3).\label{fig:gpt-4-o}}
\end{figure*}

As discussed above, the cosine similarity metric indicates the semantic closeness between the generated and reference texts. BLEU scores measure the n-gram overlap, providing insights into the textual accuracy of the models. ROUGE scores evaluate the quality of the summaries by comparing overlap in recall, precision, and F1 metrics. The results demonstrate that while all models generally show a decline in performance as the complexity of the metric increases (from cosine similarity to BLEU-4), certain models like GPT-3.5-Turbo and GPT-3.5-Turbo-1106 maintain a relatively stable performance across all metrics. In contrast, more advanced models like GPT-4 and GPT-4-1106-preview show improvements in precision and recall, reflecting their enhanced capabilities in generating accurate and contextually relevant content. These observations suggest a trade-off between computational complexity and performance, with advanced models excelling in precision-intensive tasks.  Figure \ref{fig:gpt-4-o} focuses on the GPT-4o model, comparing its performance metrics across different numbers of examples (0, 1, 2, 3). This plot provides a detailed analysis of how the number of examples impacts the model's performance across cosine similarity, BLEU scores (BLEU-1 to BLEU-4), and ROUGE scores (ROUGE Precision, Recall, and F1). \\ 

The results indicate that the inclusion of more examples generally enhances the model's performance, particularly in terms of cosine similarity and BLEU scores. For instance, with zero examples, the model's performance is lower across all metrics. However, as the number of examples increases, there is a noticeable improvement in BLEU-1 and BLEU-2 scores, which suggests better n-gram overlap and more accurate text generation. The ROUGE scores also show improvement, particularly in precision and recall, indicating that the model is better at capturing relevant content and generating precise summaries. \\

This analysis underscores the importance of providing sufficient examples to the GPT-4o model to achieve optimal performance, highlighting the model's dependency on context for generating high-quality outputs. These detailed analyses of Figures \ref{fig:all_models} and \ref{fig:gpt-4-o} provide insights into the performance dynamics of different GPT models and the impact of example quantity on the GPT-4o model, offering valuable implications for optimizing model selection and training strategies in clinical protocol generation. Metric comparison figures for the rest of the models used in this analysis are provided in the Appendix section \ref{appendix:2}. \\

Table \ref{tab:model_performance} provides a comprehensive comparison of various GPT models employed in this study with a varying number of examples provided in the prompt. The metrics evaluated include response time, input and output token counts, cosine similarity, BLEU scores (BLEU-1 to BLEU-4), and ROUGE scores (precision, recall, and F1). \\

This table offers a detailed view of each model's performance across different contexts, enabling a thorough analysis of their capabilities. Among the models, GPT-3.5-Turbo-1106 consistently demonstrates strong performance across most metrics, particularly with one example, where it achieves high cosine similarity (0.78), BLEU-1 (0.80), and BLEU-2 (0.63) scores. This indicates a high level of accuracy and relevance in the generated content. \\

On the other hand, GPT-4 shows significant improvements in more complex tasks, with notable performance in BLEU-3 (0.43) and BLEU-4 (0.34) scores when provided with two examples, highlighting its ability to handle intricate text generation tasks effectively. Despite the higher token costs, the GPT-4 model’s precision and recall metrics further underscore its advanced capabilities in producing contextually accurate and precise text. Considering the balance of performance and computational efficiency, GPT-4 stands out as the best-performing model overall, especially in scenarios requiring high accuracy and detailed text generation.\\

From a technological perspective, gpt-4 models and gpt-4o model specifically has superior performance compared to all other models, therefore it would be the logical decision to choose this model for the task of protocol generation. However, there are various other factors to take into consideration such as cost analysis for each model and limitation on the input and output text provided and retrieved from the model respectively in addition to other factors. In the next section, we delve deeper into this aspect and perform a comparative analysis across all models to provide a clearn vision into protocol generating using gpt models. 

\renewcommand{\arraystretch}{1.5}
\begin{sidewaystable*}[htbp]
\centering
\caption{Evaluation Metrics for all GPT Models and a variation number of examples provided in the prompt}
\label{tab:model_performance}
\begin{tabularx}{\linewidth}{|p{3.5cm}|l|l|l|l|X|l|l|l|l|l|l|l|l|}
    \hline
    \textbf{Model} & \textbf{\#} & \textbf{Time} & \textbf{In. Tok.} & \textbf{Out. Tok.} & \textbf{Error} & \textbf{Cos. Sim.} & \textbf{bleu1} & \textbf{bleu2} & \textbf{bleu3} & \textbf{bleu4} & \textbf{Precision} & \textbf{Recall} & \textbf{F1} \\
    \hline
    gpt-3.5-turbo & 0 & 5.06 & 373 & 214 & - & 0.57 & 0.53 & 0.26 & 0.15 & 0.09 & 0.32 & 0.13 & 0.19 \\
    gpt-3.5-turbo & 1 & 7.62 & 1188 & 246 & - & 0.73 & 0.75 & 0.53 & 0.39 & 0.30 & 0.65 & 0.28 & 0.39 \\
    gpt-3.5-turbo & 2 & 11.15 & 3554 & 392 & - & 0.60 & 0.54 & 0.25 & 0.13 & 0.08 & 0.35 & 0.20 & 0.25 \\
    gpt-3.5-turbo & 3 & 10.72 & 3577 & 346 & - & 0.78 & 0.75 & 0.53 & 0.39 & 0.29 & 0.65 & 0.37 & 0.47 \\
    gpt-3.5-turbo-1106 & 0 & 5.88 & 373 & 251 & - & 0.62 & 0.57 & 0.28 & 0.14 & 0.08 & 0.35 & 0.16 & 0.22 \\
    gpt-3.5-turbo-1106 & 1 & 5.41 & 1188 & 283 & - & 0.78 & 0.80 & 0.63 & 0.50 & 0.40 & 0.70 & 0.35 & 0.46 \\
    gpt-3.5-turbo-1106 & 2 & 4.67 & 2773 & 249 & - & 0.71 & 0.70 & 0.46 & 0.32 & 0.24 & 0.60 & 0.26 & 0.36 \\
    gpt-3.5-turbo-1106 & 3 & 5.63 & 4358 & 292 & - & 0.65 & 0.61 & 0.38 & 0.27 & 0.20 & 0.45 & 0.20 & 0.27 \\
    gpt-4 & 0 & 25.11 & 373 & 328 & - & 0.63 & 0.50 & 0.23 & 0.10 & 0.05 & 0.26 & 0.16 & 0.20 \\
    gpt-4 & 1 & 39.55 & 1969 & 499 & - & 0.71 & 0.51 & 0.24 & 0.11 & 0.05 & 0.33 & 0.30 & 0.31 \\
    gpt-4 & 2 & 38.16 & 2773 & 448 & - & 0.81 & 0.74 & 0.54 & 0.43 & 0.34 & 0.60 & 0.50 & 0.54 \\
    gpt-4 & 3 & 32.78 & 3577 & 408 & - & 0.71 & 0.61 & 0.38 & 0.24 & 0.16 & 0.47 & 0.37 & 0.41 \\
    gpt-4-1106-preview & 0 & 23.60 & 373 & 562 & - & 0.62 & 0.41 & 0.18 & 0.07 & 0.03 & 0.22 & 0.23 & 0.23 \\
    gpt-4-1106-preview & 1 & 34.25 & 1188 & 535 & - & 0.70 & 0.48 & 0.24 & 0.11 & 0.06 & 0.26 & 0.27 & 0.26 \\
    gpt-4-1106-preview & 2 & 34.37 & 1992 & 527 & - & 0.65 & 0.42 & 0.18 & 0.07 & 0.03 & 0.23 & 0.23 & 0.23 \\
    gpt-4-1106-preview & 3 & 23.63 & 2796 & 514 & - & 0.68 & 0.45 & 0.23 & 0.09 & 0.03 & 0.24 & 0.25 & 0.24 \\
    gpt-4-turbo & 0 & - & - & - & math domain error & - & - & - & - & - & - & - & - \\
    gpt-4-turbo & 1 & 24.57 & 1969 & 482 & - & 0.64 & 0.47 & 0.20 & 0.09 & 0.04 & 0.30 & 0.29 & 0.30 \\
    gpt-4-turbo & 2 & 20.43 & 2773 & 428 & - & 0.66 & 0.48 & 0.20 & 0.09 & 0.05 & 0.27 & 0.23 & 0.24 \\
    gpt-4-turbo & 3 & 35.94 & 5139 & 513 & - & 0.64 & 0.43 & 0.18 & 0.05 & 0.02 & 0.29 & 0.30 & 0.29 \\
    gpt-4o & 0 & 13.71 & 373 & 533 & - & 0.63 & 0.41 & 0.18 & 0.09 & 0.06 & 0.26 & 0.24 & 0.25 \\
    gpt-4o & 1 & 17.73 & 1969 & 671 & - & 0.72 & 0.43 & 0.21 & 0.11 & 0.06 & 0.30 & 0.37 & 0.33 \\
    gpt-4o & 2 & 10.34 & 2773 & 410 & - & 0.69 & 0.56 & 0.32 & 0.21 & 0.15 & 0.38 & 0.30 & 0.34 \\
    gpt-4o & 3 & 10.77 & 4358 & 554 & - & 0.71 & 0.50 & 0.27 & 0.15 & 0.08 & 0.35 & 0.38 & 0.36 \\
    \bottomrule
\end{tabularx}
\end{sidewaystable*}

\newpage 
\subsection{Cost Analysis}

The cost analysis experiment aimed to evaluate the economic feasibility of using different GPT models for generating clinical trial protocols. The models assessed were gpt-3.5-turbo, gpt-3.5-turbo-1106, gpt-4, gpt-4-1106-preview, gpt-4-turbo, and gpt-4o. The analysis considered generation of full protocol (aggregating each generated section) and providing multiple examples for each section. In addition, it has been assumed that each section will be generated multiple times due to human verification and evaluation and the potential need to update prompt after generation of each section.\\

All in all, it has been assumed that 50,000 tokens will be used to generate a full protocol with the assumptions described above. Finally, it has been assumed that the aggregate sections of the protocol will consume 25,000 tokens (considering the inclusion of examples and re-generation of each section as mentioned above). The calculated the input cost, output cost, total cost, and annual cost for producing 100 protocols, as shown in Table \ref{tab:cost_analysis}.\\

\renewcommand{\arraystretch}{1.5}
\setlength{\tabcolsep}{2pt} 
\begin{table*}
\caption{Forecast Cost Analysis for GPT Models with Varying Number of Examples}
\label{tab:cost_analysis}
        \begin{tabularx}{\linewidth}{p{3cm}p{2.5cm}p{1.8cm}p{1.9cm}p{1.8cm}p{2cm}}
            \toprule
            \textbf{Model} & \textbf{Sections} \newline \textbf{Generated} & \textbf{Input Cost} & \textbf{Output Cost} & \textbf{Total Cost} & \textbf{Annual Cost} \\
            \midrule
            gpt-3.5-turbo & Entire Protocol & $\$0.1K$ & $\$0.1K$ & $\$0.1K$ & $\$15.0K$ \\
            gpt-3.5-turbo-1106 & Entire Protocol & $\$0.1K$ & $\$0.1K$ & $\$0.1K$ & $\$15.0K$ \\
            gpt-4 & Entire Protocol & $\$1.1K$ & $\$1.1K$ & $\$2.2K$ & $\$225.0K$ \\
            gpt-4-1106-preview & Entire Protocol & $\$1.1K$ & $\$1.1K$ & $\$2.2K$ & $\$225.0K$ \\
            gpt-4-turbo & Entire Protocol & $\$0.4K$ & $\$0.4K$ & $\$0.8K$ & $\$75.0K$ \\
            gpt-4o & Entire Protocol & $\$0.4K$ & $\$0.4K$ & $\$0.8K$ & $\$75.0K$ \\
            \bottomrule
        \end{tabularx}
\end{table*}

\begin{figure}
\centering
\includegraphics[width=12cm]{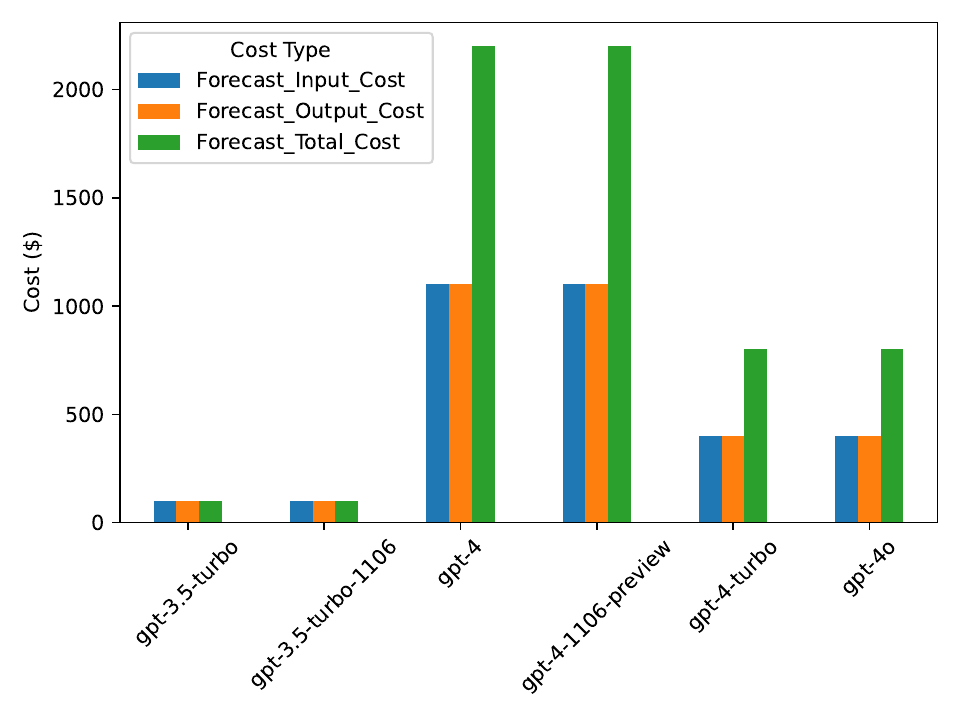}
\caption{Forecast Cost Analysis for GPT Models with Varying Number of Examples. \label{fig:cost_forecast}}
\end{figure}

Among the models, gpt-3.5-turbo and gpt-3.5-turbo-1106 exhibited the lowest total cost across all example sizes, maintaining an annual cost of \$15,000. These models demonstrated consistent cost efficiency but may lack advanced capabilities compared to newer versions. In contrast, the gpt-4 and gpt-4-1106-preview models, while providing enhanced performance, resulted in significantly higher costs, reaching an annual expenditure of \$225,000. This substantial difference highlights the trade-off between cost and model sophistication, as illustrated in Figure \ref{fig:cost_forecast}.\\

The gpt-4-turbo and gpt-4o models presented a more balanced approach, offering improved performance at a reduced cost compared to gpt-4 and gpt-4-1106-preview. With an annual cost of \$75,000, these models deliver better cost efficiency while still benefiting from the advancements in the GPT-4 series. This middle ground suggests that for organizations seeking both performance and cost-effectiveness, gpt-4-turbo and gpt-4o are preferable choices.\\

The token rates and efficiency factors used in this experiment are detailed in Table \ref{tab:token_rates_efficiency}. The token rates for input and output tokens vary significantly across the different GPT models. For instance, gpt-3.5-turbo and gpt-3.5-turbo-1106 have the lowest rates at \$0.002 per 1K tokens for both input and output, reflecting their cost efficiency but potentially lower performance capabilities. The efficiency factor has been defined as how much more efficient a given model is compared to based gpt-3.5 models and has been an entirely qualitative assessment.\\

On the other hand, the gpt-4 and gpt-4-1106-preview models have considerably higher token rates, with input tokens costing \$0.03 and output tokens costing \$0.06 per 1K tokens. These higher rates correspond to the enhanced capabilities and performance of these models. However, their efficiency factors, set at 0.75, indicate that they are 25\% more efficient in token utilization compared to the gpt-3.5-turbo models.\\

\begin{table}
\caption{Token Rates and Efficiency Factors for GPT Models}
\label{tab:token_rates_efficiency}
\centering
\begin{tabularx}{\textwidth}{p{4cm}p{3.5cm}p{3.5cm}p{3cm}}
    \toprule
    \textbf{Model} & \textbf{Input Token Rate} \newline (\$/1K tokens) & \textbf{Output Token Rate} \newline (\$/1K tokens) & \textbf{Efficiency Factor} \\
    \midrule
    gpt-3.5-turbo & \$0.002 & \$0.002 & 1.0 \\
    gpt-3.5-turbo-1106 & \$0.002 & \$0.002 & 1.0 \\
    gpt-4 & \$0.03 & \$0.06 & 1.25 \\
    gpt-4-1106-preview & \$0.03 & \$0.06 & 1.25 \\
    gpt-4-turbo & \$0.015 & \$0.03 & 1.5 \\
    gpt-4o & \$0.015 & \$0.03 & 1.5 \\
    \bottomrule
\end{tabularx}
\end{table}

As discussed above, the gpt-4-turbo and gpt-4o models strike a balance between cost and efficiency, with token rates of \$0.015 for input tokens and \$0.03 for output tokens. These models have an efficiency factor of 0.5, meaning they are 50\% more efficient than the baseline gpt-3.5-turbo models. This balance makes them a cost-effective choice for organizations seeking both performance improvements and manageable costs.\\

The implications of this cost analysis are significant for decision-making in clinical protocol generation. While advanced models like gpt-4 offer superior capabilities, their high cost may not be justifiable for all applications. Models like gpt-4-turbo and gpt-4o provide a viable alternative, balancing performance improvements with manageable costs. Organizations can thus make informed decisions based on their specific needs and budget constraints, optimizing both economic and operational efficiency in protocol development.

\section{Discussion}

The presented results underscore the substantial impact of providing examples in the prompt on enhancing the quality and coherence of text generation. Our analysis revealed that models such as GPT-4-turbo and GPT-4o significantly outperform others in almost all measured metrics. Notably, the inclusion of one to two examples proved optimal for helping the models grasp the required writing style and content for the introduction section, achieving high cosine similarity and ROUGE precision scores.\\

Our cost analysis experiment highlighted the economic efficiency of different models, revealing that GPT-4-turbo and GPT-4o offer a balanced trade-off between cost and performance. Despite the higher computational costs associated with advanced models like GPT-4, their superior output quality justifies the investment, especially in scenarios requiring high accuracy and adherence to complex regulatory standards. The annual cost analysis demonstrated substantial potential savings and resource optimization when utilizing these generative AI models for clinical protocol development.\\

The data analysis confirms the transformative potential of generative AI in clinical trial protocol authoring. GPT-4 models, in particular, demonstrated an exceptional ability to produce text that closely aligns with human expert outputs, as evidenced by high cosine similarity and ROUGE scores. These findings suggest that with adequate training and fine-tuning, generative AI can significantly reduce the time required for protocol development while maintaining or enhancing the quality and compliance of the documents produced.\\

Further examination of the figures and tables reveals detailed insights into model performance across different settings. The incremental improvements observed with an increasing number of examples highlight the importance of context-rich input data for training AI systems. This is crucial for future applications of AI in clinical research, suggesting that a more extensive dataset, possibly encompassing a broader variety of trial types and therapeutic areas, could further enhance the model's accuracy and versatility. The adaptability of GPT-4 models to the provided metadata, as demonstrated by various metrics, illustrates the potential for personalized medicine applications where protocols are tailored to specific patient populations or disease characteristics.\\

Looking ahead, the integration of AI in clinical trials could revolutionize other aspects of clinical research, such as patient recruitment and data analysis. AI's ability to learn from vast datasets and identify patterns can be leveraged to predict trial outcomes, optimize resource allocation, and enhance patient safety through more precise and individualized trial protocols. However, as we push the boundaries of AI applications in clinical settings, continuous evaluations of ethical implications, transparency, and data privacy must be prioritized to ensure that the adoption of these technologies contributes positively to patient outcomes and public trust in medical research.

\subsection{Challenges and Resolutions}

The initial use of T5 models highlighted challenges in generating long and contextually rich content. By shifting focus to the GPT-4 model and refining prompt engineering techniques, we significantly improved the output quality for the introduction section. Token limits for input or output in earlier models like GPT-3.5 were resolved in newer versions of GPT models, alleviating previous constraints. This study's scope was limited to Type II Diabetes and US-based sponsors, using clinical protocols and associated metadata to predict select protocol sections. Future research could expand to include international sponsors, various medical drugs, and different clinical trial protocol sections.

\subsection{Impact and Future Directions}

The successful implementation of this work marks a significant step forward in clinical trial design. Integrating AI-generated protocol sections into existing toolkits for trial design and study planning can lead to extensive time and financial savings, along with higher accuracy and efficiency. Future efforts may include expanding the range of automated sections, refining the model's accuracy, and exploring further integration within clinical research frameworks.

\section{Conclusion}

This work stands as a testament to the transformative potential of AI in clinical research studies. The strategic use of generative AI models like GPT-4 not only streamlined the protocol development process but also opened new avenues for innovation in clinical trial design. The research project demonstrated the immense potential of generative AI in the realm of clinical trial design. The successful integration of AI-generated protocol sections represents a significant leap forward in optimizing the efficiency and accuracy of clinical trial protocols. Future endeavors may explore broader applications of this technology, further revolutionizing the field of clinical research.

\bibliographystyle{plainnat}
\bibliography{references}

\begin{thebibliography}{35}
\providecommand{\natexlab}[1]{#1}
\providecommand{\url}[1]{\texttt{#1}}
\expandafter\ifx\csname urlstyle\endcsname\relax
  \providecommand{\doi}[1]{doi: #1}\else
  \providecommand{\doi}{doi: \begingroup \urlstyle{rm}\Url}\fi

\bibitem[cli(2024)]{clinicaltrials2024}
Clinicaltrials.gov.
\newblock \url{https://clinicaltrials.gov}, 2024.
\newblock Accessed: 2024-05-04.

\bibitem[tri(2024)]{trialtrove2024}
Trialtrove.
\newblock \url{https://www.citeline.com/trialtrove/}, 2024.
\newblock Accessed: 2024-05-04.

\bibitem[Anderson(2020)]{Anderson2020}
John Anderson.
\newblock The evolution of clinical trial design.
\newblock \emph{Journal of Clinical Research}, 12:\penalty0 45--50, 2020.

\bibitem[Brown(2019)]{Brown2019}
Sarah Brown.
\newblock The importance of clinical trials.
\newblock \emph{Medical Research Review}, 15:\penalty0 100--110, 2019.

\bibitem[Clark(2021)]{Clark2021}
Anna Clark.
\newblock Improving clinical protocols with ai.
\newblock \emph{Journal of Medical AI}, 12:\penalty0 35--45, 2021.

\bibitem[Dai et~al.(2019)Dai, Yang, Yang, Carbonell, Le, and
  Salakhutdinov]{dai2019transformer}
Zihang Dai, Zhilin Yang, Yiming Yang, Jaime Carbonell, Quoc~V Le, and Ruslan
  Salakhutdinov.
\newblock Transformer-xl: Attentive language models beyond a fixed-length
  context.
\newblock \emph{arXiv preprint arXiv:1901.02860}, 2019.

\bibitem[Dathathri et~al.(2019)Dathathri, Madotto, Lan, Hung, Frank, Molino,
  Yosinski, and Liu]{dathathri2019plug}
Sumanth Dathathri, Andrea Madotto, Janice Lan, Jane Hung, Eric Frank, Piero
  Molino, Jason Yosinski, and Rosanne Liu.
\newblock Plug and play language models: A simple approach to controlled text
  generation.
\newblock \emph{arXiv preprint arXiv:1912.02164}, 2019.

\bibitem[Davis(2018)]{Davis2018}
Richard Davis.
\newblock Challenges in clinical trial design.
\newblock \emph{Clinical Trials Journal}, 22:\penalty0 70--75, 2018.

\bibitem[Devlin et~al.(2018)Devlin, Chang, Lee, and Toutanova]{devlin2018bert}
Jacob Devlin, Ming-Wei Chang, Kenton Lee, and Kristina Toutanova.
\newblock Bert: Pre-training of deep bidirectional transformers for language
  understanding.
\newblock \emph{arXiv preprint arXiv:1810.04805}, 2018.

\bibitem[Garcia(2019)]{Garcia2019}
Robert Garcia.
\newblock Clinical trials: Phases and importance.
\newblock \emph{Journal of Clinical Studies}, 11:\penalty0 30--40, 2019.

\bibitem[Harris(2018)]{Harris2018}
Steven Harris.
\newblock Endpoints and data collection in clinical trials.
\newblock \emph{Medical Data Journal}, 8:\penalty0 40--50, 2018.

\bibitem[Jackson(2018)]{Jackson2018}
Emily Jackson.
\newblock Enhancing data quality in clinical trials.
\newblock \emph{Medical Research Innovations}, 7:\penalty0 25--35, 2018.

\bibitem[Johnson(2021)]{Johnson2021}
Emily Johnson.
\newblock Harnessing ai for clinical protocols.
\newblock \emph{Medical Innovations}, 8:\penalty0 33--40, 2021.

\bibitem[Keskar et~al.(2019)Keskar, McCann, Varshney, Xiong, and
  Socher]{keskar2019ctrl}
Nitish~Shirish Keskar, Bryan McCann, Lav~R Varshney, Caiming Xiong, and Richard
  Socher.
\newblock Ctrl: A conditional transformer language model for controllable
  generation.
\newblock \emph{arXiv preprint arXiv:1909.05858}, 2019.

\bibitem[Lee(2023)]{Lee2023}
Kevin Lee.
\newblock Generative ai in clinical research: A case study.
\newblock \emph{Clinical Innovations}, 16:\penalty0 55--65, 2023.

\bibitem[Lester et~al.(2021)Lester, Al-Rfou, and Constant]{lester2021power}
Brian Lester, Rami Al-Rfou, and Noah Constant.
\newblock The power of scale for parameter-efficient prompt tuning.
\newblock \emph{arXiv preprint arXiv:2104.08691}, 2021.

\bibitem[Lewis et~al.(2020)Lewis, Liu, Goyal, Ghazvininejad, Mohamed, Levy,
  Stoyanov, and Zettlemoyer]{lewis2020bart}
Mike Lewis, Yinhan Liu, Naman Goyal, Marjan Ghazvininejad, Abdelrahman Mohamed,
  Omer Levy, Veselin Stoyanov, and Luke Zettlemoyer.
\newblock Bart: Denoising sequence-to-sequence pre-training for natural
  language generation, translation, and comprehension.
\newblock In \emph{Proceedings of the 58th Annual Meeting of the Association
  for Computational Linguistics}, pages 7871--7880, 2020.

\bibitem[Martin(2022)]{Martin2022}
Susan Martin.
\newblock Phases of clinical trials.
\newblock \emph{Health Research Journal}, 17:\penalty0 45--55, 2022.

\bibitem[Nguyen(2020)]{Nguyen2020}
Linda Nguyen.
\newblock Study design in clinical trials.
\newblock \emph{Journal of Clinical Design}, 13:\penalty0 60--70, 2020.

\bibitem[OpenAI(2021)]{openai2021gpt35turbo}
OpenAI.
\newblock Introducing gpt-3.5 turbo.
\newblock \url{https://blog.openai.com/gpt-3-5-turbo}, 2021.
\newblock Accessed: 2024-05-04.

\bibitem[OpenAI(2023)]{openai2023gpt4}
OpenAI.
\newblock Gpt-4 model overview.
\newblock \url{https://blog.openai.com/gpt-4}, 2023.
\newblock Accessed: 2024-05-04.

\bibitem[Patel(2019)]{Patel2019}
Raj Patel.
\newblock Mitigating bias in clinical trials.
\newblock \emph{Clinical Research Methods}, 9:\penalty0 25--35, 2019.

\bibitem[Petroni et~al.(2019)Petroni, Rockt{\"a}schel, Lewis, Bakhtin, Wu,
  Miller, and Riedel]{petroni2019language}
Fabio Petroni, Tim Rockt{\"a}schel, Patrick Lewis, Anton Bakhtin, Yuxiang Wu,
  Alexander~H Miller, and Sebastian Riedel.
\newblock Language models as knowledge bases?
\newblock \emph{arXiv preprint arXiv:1909.01066}, 2019.

\bibitem[Raffel et~al.(2020)Raffel, Shazeer, Roberts, Lee, Narang, Matena,
  Zhou, Li, and Liu]{raffel2020exploring}
Colin Raffel, Noam Shazeer, Adam Roberts, Katherine Lee, Sharan Narang, Michael
  Matena, Yanqi Zhou, Wei Li, and Peter~J. Liu.
\newblock Exploring the limits of transfer learning with a unified text-to-text
  transformer.
\newblock \emph{Journal of Machine Learning Research}, 21\penalty0
  (140):\penalty0 1--67, 2020.
\newblock URL \url{http://jmlr.org/papers/v21/20-074.html}.

\bibitem[Roberts(2021)]{Roberts2021}
Jennifer Roberts.
\newblock Evaluating medical interventions through clinical trials.
\newblock \emph{Medical Research Journal}, 14:\penalty0 75--85, 2021.

\bibitem[Shahparvari et~al.(2022)Shahparvari, Hassanizadeh, Mohammadi, Kiani,
  Lau, Chhetri, and Abbasi]{shahparvari2022decision}
Shahrooz Shahparvari, Behnam Hassanizadeh, Alireza Mohammadi, Behzad Kiani,
  Kwok~Hung Lau, Prem Chhetri, and Babak Abbasi.
\newblock A decision support system for prioritised covid-19 two-dosage
  vaccination allocation and distribution.
\newblock \emph{Transportation Research Part E: Logistics and Transportation
  Review}, 159:\penalty0 102598, 2022.

\bibitem[Shahparvari et~al.(2024)Shahparvari, Hassanizadeh, Chowdhury, Lau,
  Chhetri, and Childerhouse]{shahparvari2024supply}
Shahrooz Shahparvari, Behnam Hassanizadeh, Priyabrata Chowdhury, Kwok~Hung Lau,
  Prem Chhetri, and Paul Childerhouse.
\newblock Supply chain strategies to reduce vaccine wastage for disease x: A
  covid-19 case (preprint).
\newblock 2024.

\bibitem[Smith(2020)]{Smith2020}
Michael Smith.
\newblock Patient recruitment in clinical trials.
\newblock \emph{Clinical Trials Review}, 12:\penalty0 45--55, 2020.

\bibitem[Smith(2022)]{Smith2022}
Michael Smith.
\newblock Generative ai in clinical research.
\newblock \emph{Health Tech Journal}, 5:\penalty0 25--30, 2022.

\bibitem[Song et~al.(2019)Song, Tan, Qin, Lu, and Liu]{song2019mass}
Kaitao Song, Xu~Tan, Tao Qin, Jianfeng Lu, and Tie-Yan Liu.
\newblock Mass: Masked sequence to sequence pre-training for language
  generation.
\newblock \emph{arXiv preprint arXiv:1905.02450}, 2019.

\bibitem[Thomas(2022)]{Thomas2022}
David Thomas.
\newblock Capabilities of gpt-4 in healthcare.
\newblock \emph{AI in Medicine}, 10:\penalty0 15--20, 2022.

\bibitem[Thompson(2017)]{Thompson2017}
Mary Thompson.
\newblock Challenges in clinical trial design.
\newblock \emph{Journal of Health Research}, 6:\penalty0 55--65, 2017.

\bibitem[Walker(2023)]{Walker2023}
Laura Walker.
\newblock The future of clinical trials with ai.
\newblock \emph{Journal of Medical Research}, 19:\penalty0 25--35, 2023.

\bibitem[Williams(2016)]{Williams2016}
Chris Williams.
\newblock Ethical and regulatory considerations in clinical trials.
\newblock \emph{Journal of Clinical Ethics}, 5:\penalty0 30--40, 2016.

\bibitem[Wilson(2020)]{Wilson2020}
Jessica Wilson.
\newblock Automating clinical trial protocols with ai.
\newblock \emph{Journal of AI Research}, 18:\penalty0 40--50, 2020.

\end{thebibliography}

\newpage 

\appendix
\section*{Appendix A. Study Design Generation}
\label{section-appendix-A}

In this section, we present a comparison between the generated Study Design section by the GPT-4o model and the actual Study Design section of the protocol. The generated text is highlighted in \textcolor{orange}{orange} and can be found in the following:\\\\

\textcolor{orange}{\textbf{4. Study Design}}\\

\textcolor{orange}{\textbf{4.1. Overall Design}}\\

\textcolor{orange}{Study I8H-MC-BDCM (BDCM) is a multicenter, randomized, open-label, parallel, comparator-controlled Phase 2 study designed to evaluate the efficacy and safety of LY3209590 as compared to insulin degludec in patients with Type 2 Diabetes Mellitus (T2DM) who have been treated with a stable dose of metformin (alone or in combination with a stable dose of a DPPIV inhibitor and/or a SGLT2 inhibitor) for at least 3 months prior to screening.}\\

\textcolor{orange}{The study will consist of 3 periods:}\\
\textcolor{orange}{- \textbf{Study Period 1:} Screening and lead-in period, approximately 2 weeks}\\
\textcolor{orange}{- \textbf{Study Period 2:} Treatment period, 26 weeks}\\
\textcolor{orange}{- \textbf{Study Period 3:} Safety follow-up period, 5 weeks}\\

\textcolor{orange}{The study schema is presented in Section 1.2. Study governance considerations are described in detail in Section 10.1.}\\

\textcolor{orange}{\textbf{4.1.1. Study Visits}}\\

\textcolor{orange}{\textbf{Study Period 1: Screening and Lead-in}}\\

\textcolor{orange}{The purpose of procedures at screening is to establish eligibility for inclusion in the study (see Sections 5.1 and 5.2). During this period (upon signing the Informed Consent Form [ICF]), study participants will be trained on disease monitoring and management procedures, study diaries, and study protocols. A LibrePro sensor for flash glucose monitoring (FGM) will be inserted and activated at Visit 1 for baseline FGM assessment for eligible participants. Electronic participant diaries and participant paper note sheets will be provided at Visit 2 as specified in the Schedule of Activities (Section 1.3).}\\

\textcolor{orange}{\textbf{Study Period 2: Treatment Period}}\\

\textcolor{orange}{\textbf{Randomization (Week 0 [Visit 3])}}\\

\textcolor{orange}{Participants who continue to meet the eligibility criteria will be randomized into one of the treatment groups. All measures, including a urine pregnancy test, are to be done prior to the injection of the assigned investigational product (IP). The FGM sensor should be removed prior to study treatment, with data uploaded by the site. The first dose of IP will be injected by study personnel.}\\

\textcolor{orange}{\textbf{Treatment}}\\

\textcolor{orange}{Following randomization, participants will enter a 26-week treatment period. For patients randomized to the LY3209590 group, IP will be administered according to Algorithm 2. Initially, site personnel will administer IP and provide training on self-administration. The training should include information on reconstitution of IP, injection site locations, technique, and signs and symptoms of local adverse reactions.}\\

\textcolor{orange}{For patients randomized to the insulin degludec arm, insulin will be self-administered daily, starting at 10 IU/day according to a modified Riddle algorithm (Section 10.8). Concomitant metformin (with or without a DPPIV inhibitor and/or SGLT2 inhibitor) usage will continue throughout the treatment period, barring dose adjustments or discontinuation due to country-specific labels or study protocols.}\\

\textcolor{orange}{\textbf{Self-Monitoring of Blood Glucose (SMBG)}}\\

\textcolor{orange}{All patients will be provided with blood glucose meters to record measurements in the provided electronic diary. FG measurements should be performed daily, with two 6-point SMBG profiles on nonconsecutive days in the week prior to required visits noted in the Schedule of Activities (Section 1.3).}\\

\textcolor{orange}{\textbf{Flash Glucose Monitoring (FGM)}}\\

\textcolor{orange}{Participants will undergo 3 FGM sessions using the LibrePro System (Abbott) before randomization, at Visit 15, and Visit 20. Patients will remain blinded to these assessments until the study concludes.}\\

\textcolor{orange}{\textbf{Study Period 3: End-of-Treatment and Safety Follow-up}}\\

\textcolor{orange}{All randomized participants will undergo a comprehensive efficacy and safety evaluation at Visit 20 and a safety follow-up visit 5 weeks after the last dose of LY3209590 or insulin degludec. Participants will continue SMBG and may receive an appropriate glucose-lowering regimen as needed. Participants are required to return used and unused IP to the investigative site.}\\

\textcolor{orange}{Participants who discontinue IP early will undergo similar comprehensive assessments at approximately 1 week and 5 weeks after the last dose of LY3209590. If a participant is unwilling to return for follow-up, they will perform an early termination visit as their final study visit.}\\

\textcolor{orange}{\textbf{4.2. Scientific Rationale for Study Design}}\\

\textcolor{orange}{This study aims to evaluate LY3209590's effects on glycemic control, hypoglycemia, and weight gain in patients with T2DM when compared with daily basal insulin (insulin degludec). The 26-week duration is adequate to assess efficacy and safety of LY3209590.}\\

\textcolor{orange}{\textbf{4.3. Justification for Dose Selection}}\\

\textcolor{orange}{The LY3209590 dosing regimen involves an initial dose, followed by weekly adjustments. The initial dose considers baseline fasting glucose, body weight, and available data. Dose adjustments are based on prior fasting glucose and hypoglycemic events. The insulin degludec starting dose is 10 IU/day, titrated according to fasting glucose levels and safety concerns, using a modified Riddle algorithm.}\\

\textcolor{orange}{Initiation of LY3209590 is designed to achieve therapeutic goals within 12 weeks, with individual dose adjustments to optimize response. Dosing guidance is provided based on fasting glucose and hypoglycemia (Section 10.8). Safety will be closely monitored, particularly during early dose titration. Modifications to the dosing algorithm will be guided by emerging data from the sponsor.}\\

\textcolor{orange}{\textbf{4.4. End of Study Definition}}\\

\textcolor{orange}{The end of the trial is defined as the date of the last visit or last scheduled procedure shown in the Schedule of Activities (Section 1.3) for the last patient.}\\

And the reference text for the Study Design from the actual protocol corresponding to the same metadata information as the input to the gpt-4o model is highlighted in \textcolor{teal}{teal} and can be found in the following:\\\\

\textcolor{teal}{\textbf{5. Study Design}}\\

\textcolor{teal}{\textbf{5.1. Overall Design}}\\

\textcolor{teal}{Study I8H-MC-BDCM (BDCM) is a multicenter, randomized, open-label, comparator-controlled Phase 2 study with 3 study periods. A total of 536 study participants with T2DM, previously treated with basal insulin will be screened. Approximately 375 study participants will be randomized in a 1:1:1 ratio to LY3209590 in one of two dosing algorithms (Appendix 7) or insulin degludec (modified Riddle algorithm, Section 7.2.1.2). The study is designed to evaluate the efficacy and safety of LY3209590 compared with insulin degludec.}\\

\textcolor{teal}{The study will consist of 3 periods: 2-week screening, 32-week treatment period, and 5-week safety follow-up period. Study governance considerations are described in detail in Appendix 3. Figure BDCM.2 illustrates the study design.}\\

\textcolor{teal}{\textbf{5.1.1. Study Visits}}\\

\textcolor{teal}{\textbf{Study Period 1: Screening and Lead-in}}\\

\textcolor{teal}{The purpose of procedures at screening is to establish eligibility for inclusion in the study (see Sections 6.1 and 6.2). During this period, study participants will be trained on disease monitoring and disease management procedures, study diaries, and study procedures. Visit-specific subject diaries will be dispensed at Visit 2 and as specified in the Schedule of Activities (Section 2) for future visits.}\\

\textcolor{teal}{During the lead-in period, study participants should continue the same formulation and dose of their previously used insulin therapy, as well as allowable oral antidiabetic medications (OAMs - sulfonylureas [SU] and meglitinides, dipeptidyl peptidase-4 [DPP-IV] inhibitors, sodium-glucose cotransporter 2 [SGLT-2] inhibitors, biguanides, and alpha-glucosidase inhibitors) in order to allow reliable assessment of hemoglobin A1c (HbA1c) at randomization (see Section 7.7.2).}\\

\textcolor{teal}{\textbf{Continuing Glucose Monitoring:}}\\

\textcolor{teal}{A standard system will be used for continuing glucose monitoring (CGM) in an unblinded mode. The study participants will wear this device beginning on Day 0 at randomization, as shown in the Schedule of Activities (Section 2). In addition, all study participants will be allowed to use their blood glucose meters for additional blood glucose testing or for calibration of the CGM system if deemed necessary, or for taking self-monitored plasma glucose (SMPG) measurements during the outpatient period.}\\

\textcolor{teal}{At the randomization visit (Visit 3), study participants will be trained on the use of the CGM device, CGM sensor replacement, interpretation of CGM-based blood glucose values and alarms, and the requirements for CGM. For the first CGM session, study participants meeting all study-entry criteria will have the CGM sensor inserted as part of the Visit 3 activities.}\\

\textcolor{teal}{\textbf{Study Period 2: Treatment Period}}\\

\textcolor{teal}{\textbf{Randomization (Week 0 [Visit 3])}}\\

\textcolor{teal}{Study participants who continue to be eligible for the study will be randomized to one of the treatment arms. All measures to be performed at Visit 3 (see Schedule of Activities, Section 2) should be done prior to injection of the study participant’s assigned investigational product (IP) to ensure that appropriate baseline measurements are obtained. Urine and serum pregnancy tests will be done before IP administration. The study participant will begin IP, if the urine pregnancy test is negative. If the result of the serum pregnancy test is positive, the study participant will be discontinued from the study (see Section 8.2).}\\

\textcolor{teal}{Study personnel will inject the first dose of IP at the study site. The PK samples and ECGs at randomization must be collected as described in the Schedule of Activities (Section 2).}\\

\textcolor{teal}{Following randomization, study participants will participate in a 32-week treatment period. During Weeks 1 to 8, site personnel will administer IP, and study participants will receive education and training on how to self-administer IP. The training should include information on reconstitution of IP, appropriate injection site locations, injection technique, and the signs and symptoms of local adverse reactions, should that occur. During Weeks 9 to 12, IP will be reconstituted and administered at the site by the study participant, under supervision of trained site personnel to assure that the study participant is capable of self-administration. During Weeks 13 to 31, IP will be self-administered by the study participant at home, or optionally can be administered once weekly by site personnel. Information on self-injection can be reviewed as necessary throughout the study.}\\

\textcolor{teal}{LY3209590 will be administered as one of two dose individualization algorithms (Appendix 7). Insulin degludec will be self-administered daily by study participants after a training and first administration under site personnel supervision on Day 1 according to a modified Riddle algorithm (Section 7.2.1.2).}\\

\textcolor{teal}{\textbf{General considerations}}\\

\textcolor{teal}{Study procedures will be performed as listed in the Schedule of Activities (Section 2). In order to allow timely sampling for PK assessments, visits for collection of the samples will be scheduled within the required time windows provided in the Schedule of Activities (Section 2).}\\

\textcolor{teal}{Study participants will continue to use concomitant OAMs throughout the treatment period: discontinuation or changes to dose are not permitted, except in situations where dose adjustment or complete discontinuation is required per country-specific label or when allowed per study protocol (for further details see Section 7.7.2).}\\

\textcolor{teal}{Study participants will be instructed to document fasting plasma glucose (FPG) each day by using the values displayed on their CGM device after wakening or using a finger-stick glucose measurement. In addition, two 6-point SMPG profiles (prior to and 2 hours after the morning, midday, and evening meals) should be documented on non-consecutive days in the week prior to the required visits noted in the Schedule of Activities using CGM-based glucose values, (Section 2).}\\

\textcolor{teal}{Study participants who develop severe, persistent hyperglycemia based on prespecified thresholds (see Section 7.8.2.3) will receive a new glucose-lowering intervention (or rescue therapy) based on clinical judgment of the investigator (see Table BDCM.4). Study participants who need rescue therapy will continue on IP in the trial until they complete all study visits.}\\

\textcolor{teal}{\textbf{Study Period 3: End-of-Treatment and Safety Follow-Up, Visits 21, 801, 802, and 803}}\\

\textcolor{teal}{All randomized study participants should have a comprehensive efficacy and safety evaluation approximately 1 week after the last dose of IP and a safety follow-up visit approximately 6 weeks after the last dose of IP. During the safety follow-up study period, study participants will continue CGM and may be treated with an appropriate glucose-lowering regimen (see Table BDCM.4) if necessary. Study participants will also be required to return any remaining study diaries and used or unused IP to the investigative site.}\\

\textcolor{teal}{Study participants who complete the treatment period will have a comprehensive End-of-Treatment efficacy and safety assessment (Week 32; Visit 21) approximately 1 week after the last dose of IP and a safety follow-up assessment (Visit 801) approximately 6 weeks after the last dose of IP and 5 weeks after their End-of-Treatment assessment (Visit 801). During the safety follow-up study period, study participants will continue CGM and may be treated with an appropriate glucose-lowering regimen (see Table BDCM.4) if necessary. Study participants will also be required to return any remaining study diaries and used or unused IP to the investigative site.}\\

\textcolor{teal}{Study participants who discontinue IP prior to completion of the study for any reason should also have a comprehensive End-of-Treatment efficacy and safety assessment (Visit 802) approximately 1 week after the last dose of IP or as soon as reasonably possible thereafter, and should have a safety follow-up assessment (Visit 803) approximately 6 weeks after the last dose of IP. Study activities at Visits 802 and 803 are identical to Week 32 (Visit 21) and Visit 801, respectively.}\\

\textcolor{teal}{Study participants who discontinue IP prior to study completion will be encouraged to remain in the study and to complete any scheduled study procedures that occur after their safety follow-up (Visit 803). Study participants remaining in the study will receive an appropriate rescue glucose-lowering regimen. Table BDCM.4 provides details on the use of glucose-lowering medications for rescue and for treatment of acute conditions.}\\

\textcolor{teal}{Study participants who discontinue IP for any reason and are unwilling to return for a safety follow-up visit, will be asked to perform an early termination (ET) visit as their final study visit. At this visit, study participants will perform procedures listed in the Schedule of Activities (Section 2).}\\

\textcolor{teal}{\textbf{5.2. Number of Participants}}\\

\textcolor{teal}{Approximately 536 participants will be screened to achieve 375 randomized and 300 evaluable participants for an estimated total of 100 evaluable participants per treatment group.}\\

\textcolor{teal}{\textbf{5.3. End of Study Definition}}\\

\textcolor{teal}{End of the study is the date of the last visit or last scheduled procedure shown in the Schedule of Activities (Section 2) for the last study participant.}\\

\textcolor{teal}{\textbf{5.4. Scientific Rationale for Study Design}}\\

\textcolor{teal}{This study will evaluate 2 dosing regimens of LY3209590 in study participants with T2DM who are already treated with a daily basal insulin. Study participants will take their last dose of daily basal insulin the day prior to randomization. Treatment will be initiated after randomization with either LY3209590 or insulin degludec as an active comparator. As LY3209590 has a long half-life, a weekly dosing regimen is expected to reach steady-state of PK in over CCI of treatment. Therefore, the withdrawal of a daily basal insulin would result in loss of glycemic control for several weeks after randomization. Therefore, a loading-dose regimen approach, by which an exposure to achieve comparable glycemic control to prior basal insulin dose can be achieved CCI, followed by CCI starting at Week 2 to maintain steady-state exposure. Subsequently, dose adjustments can be made in the initial dose individualization phase of the study. See Appendix 7 and section 7.2.1.2 for dosing and administration details. The 32-week treatment duration is planned as the slow dose escalation approach is necessary for LY3209590 to achieve near comparable glycemic efficacy to prior daily basal insulin.}\\

\textcolor{teal}{Insulin degludec is an unblinded active comparator in this study and will be used to compare the effects of LY3209590 on glycemic control, hypoglycemia, and weight gain with a daily basal insulin.}\\

\textcolor{teal}{\textbf{5.5. Justification for Dose}}\\

\textcolor{teal}{Following a once-weekly SC administration of LY3209590, the time to reach steady-state glucose level is estimated between 4 to 10 weeks based on the long half-life of LY3209590. Therefore, weekly dose and gradual dose adjustment is recommended for LY3209590 dosing regimen.}\\

\textcolor{teal}{SAD and MAD data. Dose adjustments are based on average of prior CCI fasting glucose and hypoglycemia events. Alterations to the doses recommended by these dose adjustment algorithms are also under discretion of the investigator and will take hypoglycemia and other study participant safety concerns into account.}\\

\textcolor{teal}{The starting dose for insulin degludec is the same dose as the basal insulin that the study participant used previously before entering the study. The study participant will start a titration phase based on fasting glucose achievements under this therapy and the presence of hypoglycemia or other safety concerns. The titration will be performed each week. The dose adjustment algorithm is adapted from the well-established Riddle algorithm (Section 7.2.1.2) under discretion of the investigator.}\\

\textcolor{teal}{Guidance is provided for the two algorithms used in dosing LY3209590. There is also guidance for conversion of prior basal insulin dose to LY3209590 loading dose, as well as dosing suggestions for investigators based on fasting glucose and hypoglycemia of the study participant (Appendix 7).}\\

\textcolor{teal}{Safety of study participants will be closely monitored during the early stages of dose titration to determine whether adjustments to the conversion and dose adjustment algorithms are needed. As additional data emerges, guidance from the Sponsor on these dosing algorithms may be modified (Appendix 7).}\\

By examining these comparisons, we can evaluate the fidelity of the GPT-4o model's output against the actual protocol. The corresponding metadata information is provided to give context to each section, ensuring a clear understanding of the generated and reference texts. The goal is to assess the capability of the GPT-4o model in accurately replicating complex clinical documents, which is critical for its application in clinical trial protocol generation.


\newpage 

\section*{Appendix B. Metric Comparison}
\label{section-appendix-B}

The appendix section provides a detailed metric comparison for various GPT models evaluated in this study, excluding the GPT-4o model. The models analyzed include GPT-3.5-Turbo, GPT-3.5-Turbo-1106, GPT-4, GPT-4-1106-preview, and GPT-4-turbo. Each model's performance metrics were assessed with varying numbers of examples (0, 1, 2, 3). \\

\begin{figure}
\centering
\includegraphics[width=13cm]{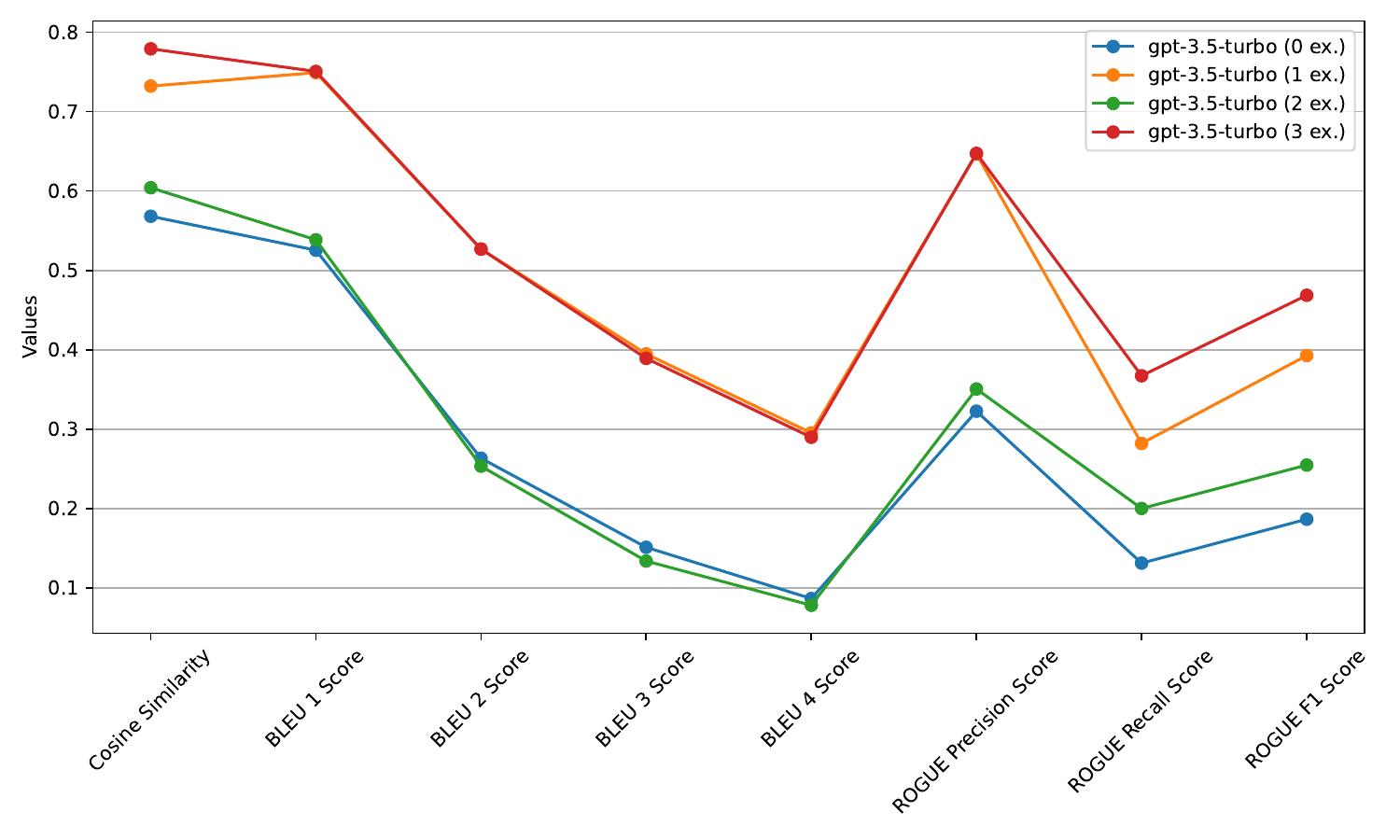}
\caption{Metric comparison for GPT-3.5-Turbo model with varying number of examples (i.e. 0, 1, 2, 3).\label{fig:gpt-3.5-turbo}}
\end{figure}

\begin{figure}
\centering
\includegraphics[width=13cm]{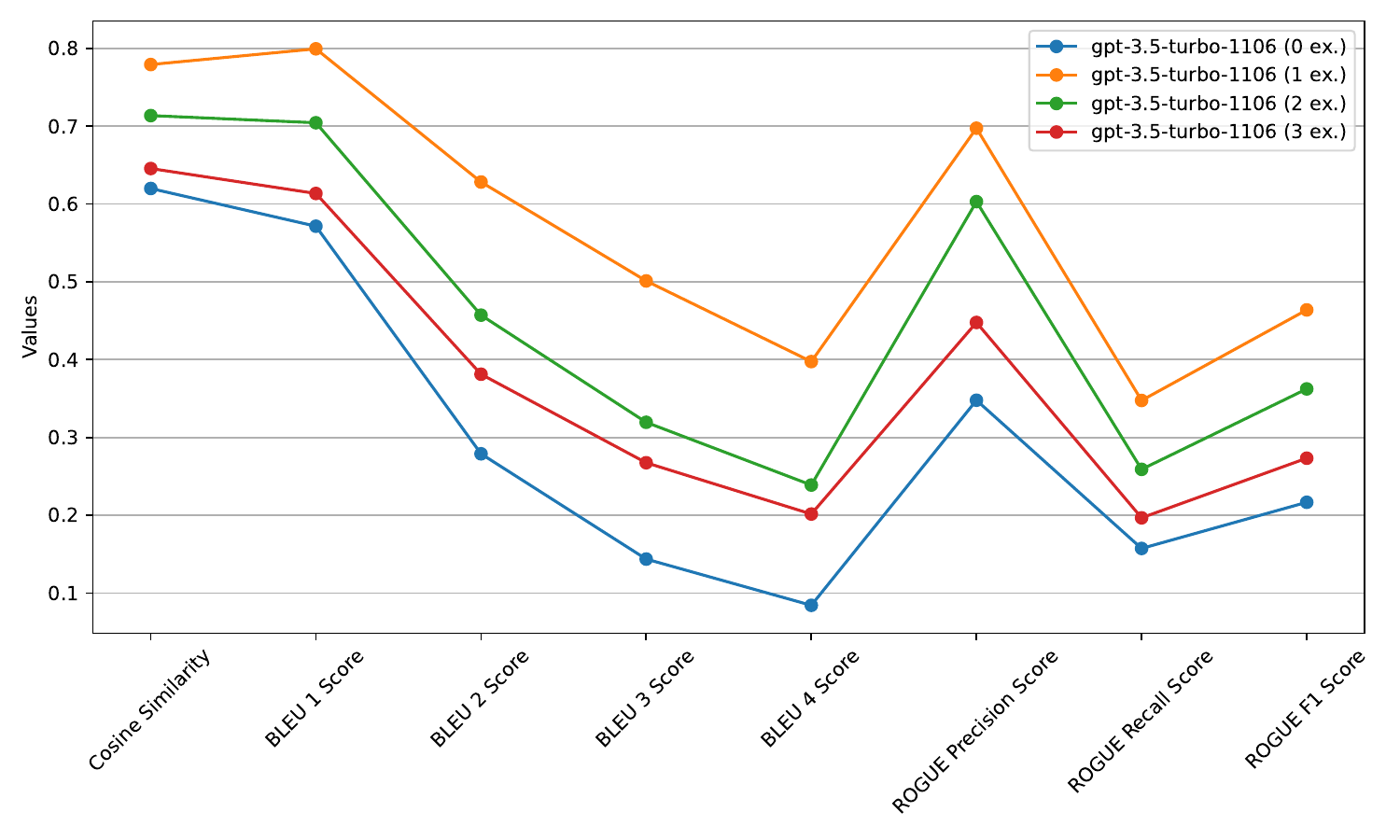}
\caption{Metric comparison for GPT-3.5-Turbo-1106 model with varying number of examples (i.e. 0, 1, 2, 3).\label{fig:gpt-3.5-turbo-1106}}
\end{figure}

\begin{figure}
\centering
\includegraphics[width=13cm]{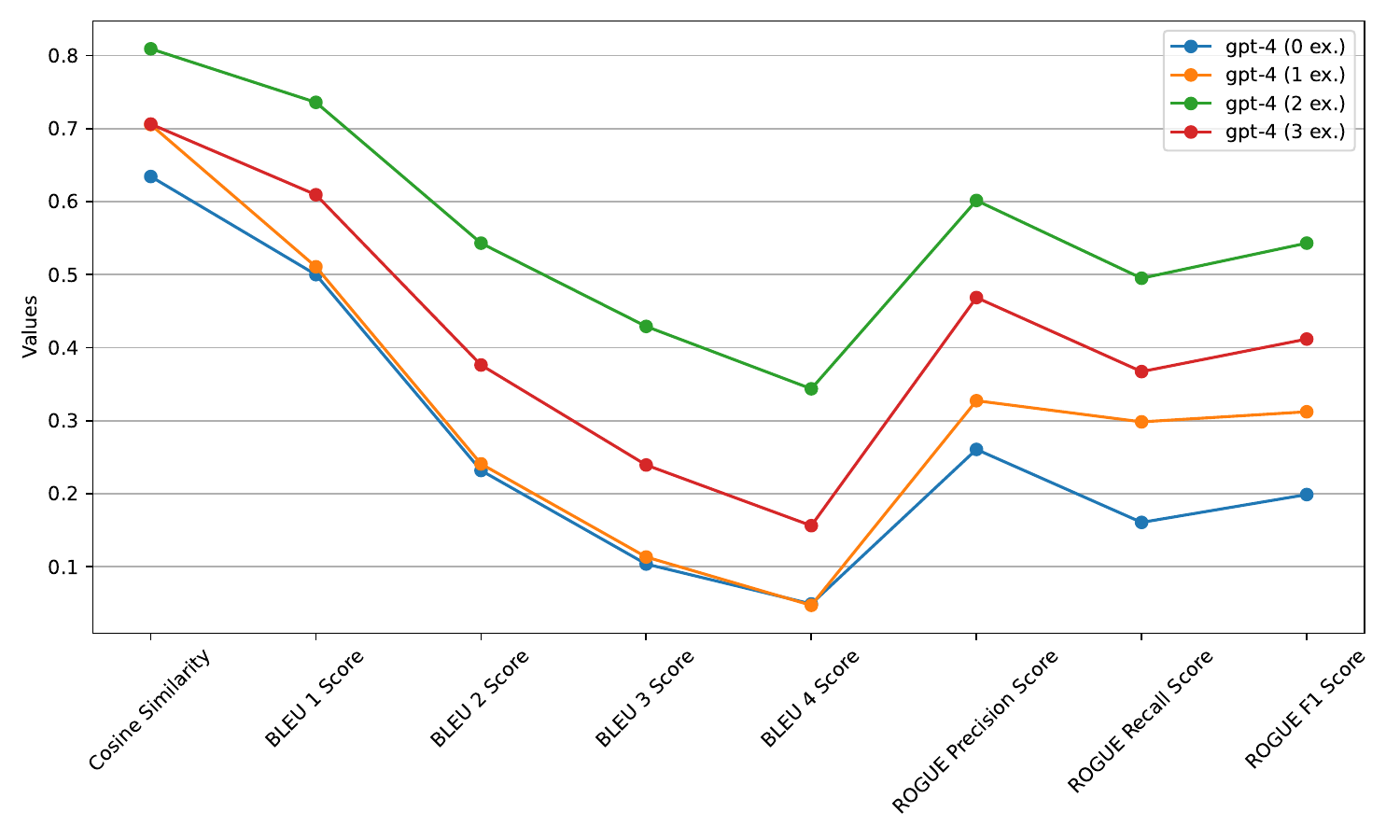}
\caption{Metric comparison for GPT-4 model with varying number of examples (i.e. 0, 1, 2, 3).\label{fig:gpt-4}}
\end{figure}

\begin{figure}
\centering
\includegraphics[width=13cm]{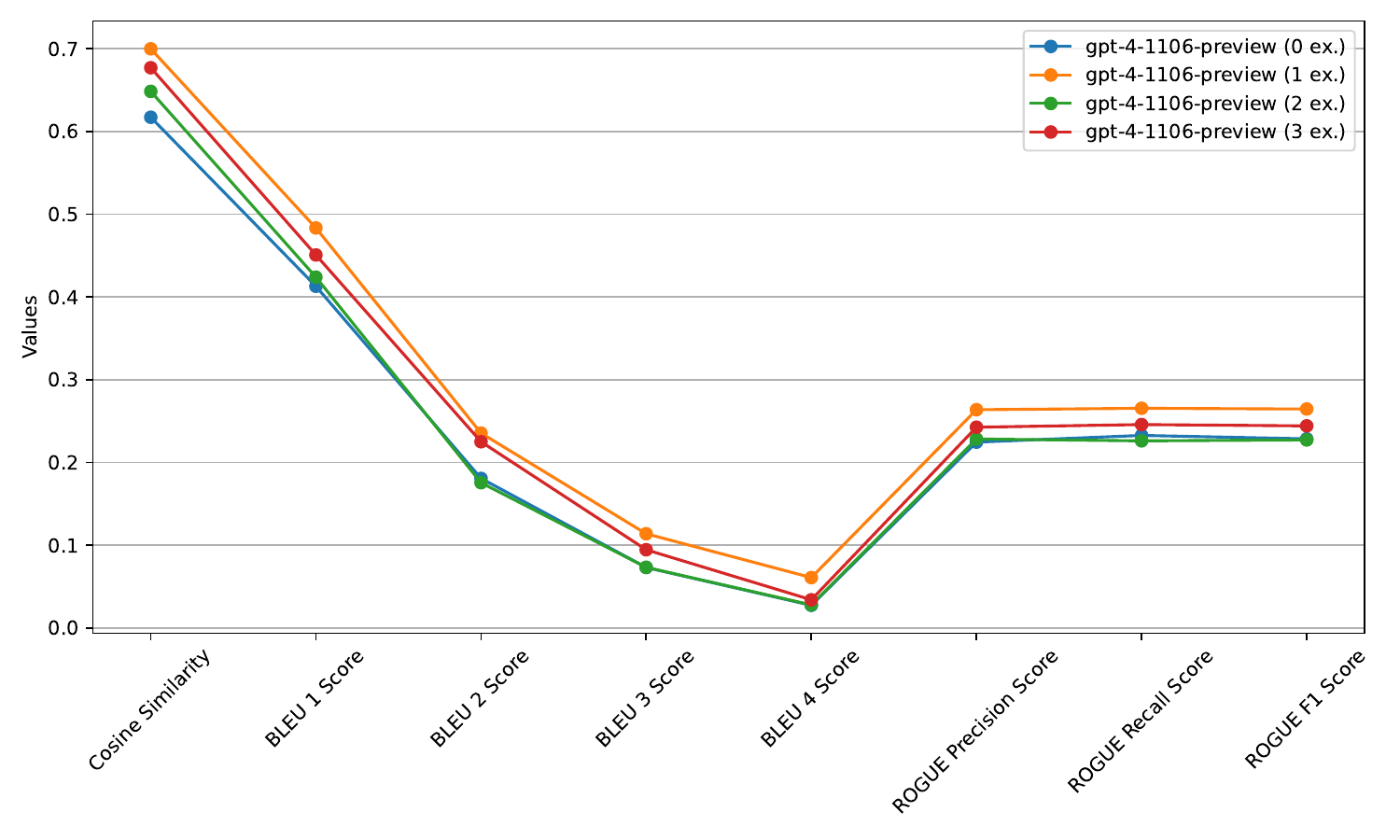}
\caption{Metric comparison for GPT-4-1106-preview model with varying number of examples (i.e. 0, 1, 2, 3).\label{fig:gpt-4-1106}}
\end{figure}

\begin{figure}
\centering
\includegraphics[width=13cm]{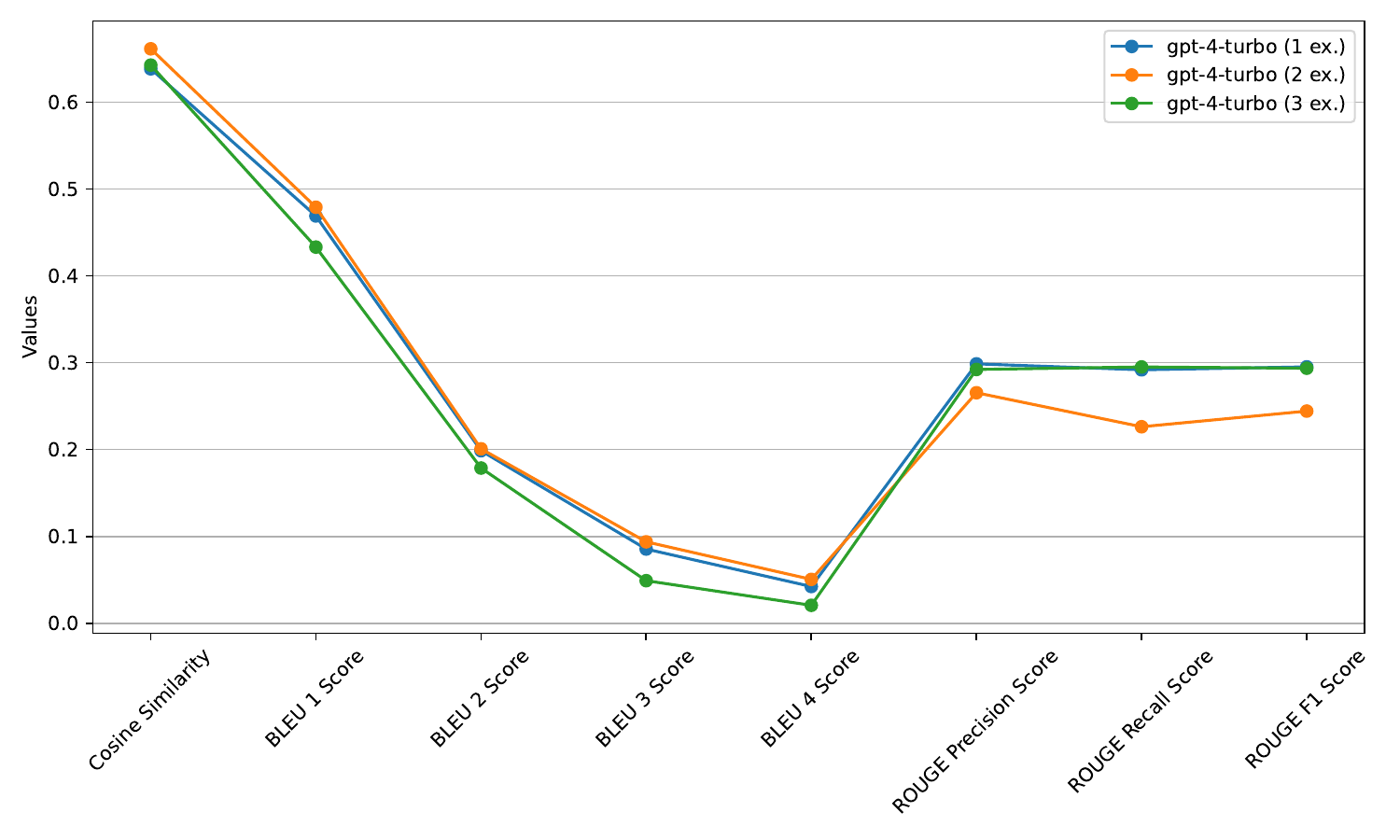}
\caption{Metric comparison for GPT-4-turbo model with varying number of examples (i.e. 1, 2, 3).\label{fig:gpt-4-turbo}}
\end{figure}

The plots illustrate the metric comparisons for each model, highlighting their efficiency and effectiveness across different scenarios. Specifically, Figure \ref{fig:gpt-3.5-turbo} shows the metric comparison for the GPT-3.5-Turbo model, Figure \ref{fig:gpt-3.5-turbo-1106} for the GPT-3.5-Turbo-1106 model, Figure \ref{fig:gpt-4} for the GPT-4 model, Figure \ref{fig:gpt-4-1106} for the GPT-4-1106-preview model, and Figure \ref{fig:gpt-4-turbo} for the GPT-4-turbo model. These plots provide valuable insights into the performance dynamics of each model, allowing for a comprehensive evaluation of their capabilities.


\end{document}